\documentclass[final,5p,times,twocolumn]{elsarticle}

\biboptions{comma,sort&compress}

\usepackage{graphicx}
\usepackage{amsmath}
\usepackage{amssymb}
\usepackage{here}
\usepackage{subfigure}
\usepackage{color}
\usepackage{times}
\usepackage{latexsym}
\usepackage{epsfig}
\usepackage{float}

\newcommand{\bcl}{\begin{center}}
\newcommand{\ecl}{\end{center}}

\usepackage{ecrc}

\volume{00}
\firstpage{1}

\journalname{Journal of Subatomic Particles and Cosmology}

\runauth{R. Perez-Ramos et al.}

\jid{jspc}

\jnltitlelogo{Journal of Subatomic Particles and Cosmology}

\begin{document}

\begin{frontmatter}


\title{The dead-cone effect in charm, bottom and top quark jets\\
in angular and momentum space\tnoteref{tnote1}}

\tnotetext[tnote1]{Talk given by R. Perez-Ramos at the 29th
International Conference in Quantum Chromodynamics (QCD 26),
20--24 July 2026, Montpellier, France.}

\author[label1,label2]{Redamy Perez-Ramos}

\author[label3]{Stefan Kluth}

\author[label3]{Wolfgang Ochs}

\address[label1]{DRII-IPSA, Bis, 63 Boulevard de Brandebourg,
94200 Ivry-sur-Seine, France}

\address[label2]{Laboratoire de Physique Th\'eorique et Hautes Energies
(LPTHE), UMR 7589, Sorbonne Universit\'e et CNRS,
4 place Jussieu, 75252 Paris Cedex 05, France}

\address[label3]{Max-Planck-Institut f\"ur Physik,
Boltzmannstrasse 8, 85748 Garching, Germany}


\pagestyle{myheadings}
\markright{ }

\begin{abstract}

The dead-cone effect provides a direct manifestation of the mass
dependence of QCD radiation from heavy quarks. We investigate this
phenomenon over a broad range of quark masses, from charm and bottom
to top quarks. Parton-level two-dimensional angular distributions are
used to geometrically illustrate  the depletion of radiation around the
heavy-quark direction and its characteristic angular scale. These
geometrical observables are complemented by the single-particle
inclusive momentum distribution, which is established as an observable 
well suited for 
studying the dead-cone effect as well.
For charm and bottom quark jets, the momentum distributions are
reconstructed from LEP data at the $Z$ pole. The strong suppression of
high-momentum particles is described quantitatively within the Modified
Leading Logarithmic Approximation of perturbative QCD. The analysis is
then extended to $\sqrt{s}=1$~TeV, where the evolution from charm and
bottom to the top-quark mass scale is investigated within a common
kinematic Monte Carlo framework. A
hypothetical stable-top sample provides a clean reference for the
primary top-quark radiation pattern. In the physical case, the decay
$t\to bW$ introduces coherent radiation associated with the
$\widehat{tb}$ colour dipole, which obscures the primary dead cone.
By exploiting the dependence of this contribution on the $b$-quark
emission angle $\Theta_b$, the primary parton momentum spectrum associated with
the top-quark jet can be reconstructed through an extrapolation towards
the forward $b$-quark configuration $\Theta_b\to0$~\cite{Kluth:2025xpj}. In contrast,
the angular distribution of gluons and partons at small angles does not converge towards the
stable-top dead-cone pattern in this limit $\Theta_b\to0$, owing to the 
strong radiation from the b-quark in the top-quark decay. 
This difference emphasizes the particular robustness
of the inclusive momentum spectrum as a quantitative observable for the
top-quark dead-cone effect. The combined charm, bottom, and top results
provide a unified picture of mass-dependent QCD radiation in momentum
space.

\end{abstract}



\end{frontmatter}

\section{Introduction}
\label{sec:introduction}

The suppression of gluon radiation in the forward direction of an
energetic heavy quark is one of the characteristic mass effects of
perturbative Quantum Chromodynamics (QCD). For a heavy quark $Q$ of
mass $M_Q$ and energy $E_Q$, gluon emission is suppressed at angular
scales below the characteristic angle
\begin{equation}
    \Theta_0 \simeq \frac{M_Q}{E_Q},
    \label{eq:Theta_0}
\end{equation}
giving rise to the so-called dead-cone effect
\cite{Dokshitzer:1991fc,Dokshitzer:1991fd}. At angles
$\Theta\gg\Theta_0$, the influence of the quark mass becomes
negligible and the radiation pattern approaches that of a light-quark
jet, whereas for $\Theta\lesssim\Theta_0$ gluon radiation is
progressively depleted.
The angular radiation pattern provides the most direct geometrical
picture of the dead cone. Two-dimensional angular distributions around
the heavy-quark direction can make the depleted forward region visible
and identify the characteristic scale $\Theta_0$. Such distributions
are therefore particularly useful for illustrating the origin of the
effect and its dependence on the heavy-quark mass. The principal
quantitative observable considered in this work, however, is the
single-particle inclusive distribution in momentum space, or
equivalently the fragmentation function of the heavy-quark jet.
The connection between these descriptions follows from the
development of the QCD cascade. Through coherent branching and angular
ordering~\cite{Ermolaev:1981cm,Mueller:1981ex,Dokshitzer1991Basics},
the suppression of radiation at small angles propagates
through the QCD cascade and modifies the momentum distribution of
particles produced in the jet. The dead cone consequently leads to a
characteristic depletion of the high-momentum part of the heavy-quark
fragmentation function. Introducing the momentum variables
\begin{equation}
    x_p=\frac{2p}{W},\qquad
    \xi_p=\ln\frac{1}{x_p},
    \label{eq:xp_xip}
\end{equation}
with the total energy $W=\sqrt{s}$, this effect appears as a suppression
at small $\xi_p$, corresponding to large particle momentum fractions $x_p$.
This momentum-space manifestation of the dead cone has been studied
using identified charm- and bottom-quark events in $e^+e^-$
annihilation at the $Z$ pole, $\sqrt{s}=91.2$~GeV
\cite{Kluth:2023umf}. After correcting the measured particle spectra
for contributions from decays of the primary heavy hadrons, genuine
charm- and bottom-quark fragmentation functions were reconstructed
from LEP data~\cite{DELPHI:1998cgx,OPAL:1998arz}. Relative to
light-quark jets, both distributions show a strong suppression at
large momentum fraction, with a characteristic flavour dependence
associated with the different heavy-quark masses.

Of particular importance is that this suppression admits a
quantitative interpretation within the Modified Leading Logarithmic
Approximation (MLLA) of perturbative QCD
\cite{Dokshitzer:1991fc,Dokshitzer:1991fd}. In this framework, the
radiation missing from the heavy-quark jet is related to a light-quark
fragmentation spectrum evaluated at a lower scale determined by the
heavy-quark mass. The comparison with the LEP measurements provides a
successful description of the observed suppression for charm and
bottom over the central kinematic region~\cite{Kluth:2023umf}.

Experimentally, the dead-cone effect was observed directly for the
first time through an angular analysis of charm-quark jets by the ALICE
Collaboration~\cite{ALICE:2021aqk}. It was subsequently demonstrated
for charm quarks and, for the first time, for bottom quarks in our
momentum-space analysis~\cite{Kluth:2023umf}. More recently, the dead-cone effect in bottom-quark jets has also been
investigated at the LHC by studying the angular separation between
reconstructed subjets on the Lund jet plane in comparison with Monte
Carlo (MC) simulations and other methods~\cite{LHCb:2025mcq,LHCb:2026mbl,CMS:2026lkj,CMS:2025afq}.


In the first part of the present work, we revisit the charm- and
bottom-quark case at the $Z$ pole and complement this momentum-space
analysis with a parton-level illustration of the angular radiation
pattern using the \textsc{Pythia} Monte Carlo event
generator~\cite{Bierlich:2022pfr}. A two-dimensional parton-level
distribution is used to illustrate the radiation geometry in coordinates
normalized to the characteristic scale $\Theta_0=M_Q/E_Q$.

The second part extends the analysis to $\sqrt{s}=1$~TeV, where the
heavy-quark mass hierarchy is considered within a common kinematic
framework and the angular analysis is then focused on the top-quark
case. For the top quark we first introduce a
hypothetical stable-top reference which isolates the primary
mass-dependent radiation pattern. We then address the physical
$t\to bW$ decay, whose additional coherent radiation obscures the
primary top dead cone, and we develop a procedure to reconstruct the corresponding primary parton momentum spectrum.
The aim of the present study is therefore to connect the geometrical
and momentum-space manifestations of the dead-cone effect from charm
and bottom to top quarks. The angular distributions make the
characteristic scale $\Theta_0=M_Q/E_Q$ and the corresponding
radiation geometry directly visible, while the single-particle
inclusive momentum distribution provides the principal quantitative
observable through which the mass-dependent suppression and its
perturbative-QCD description are tested.

\section{Theoretical prerequisites}
\label{sec:theory}

\subsection{Soft-gluon radiation from a heavy quark}
\label{subsec:deadcone}

We first recall the perturbative-QCD origin of the dead-cone effect.
Consider an energetic heavy quark $Q$ of mass $M_Q$ and energy $E_Q$
emitting a soft gluon of energy $\omega$ at an angle $\Theta$ with
respect to the primary-quark direction. In the soft and small-angle
approximation, the differential gluon-emission cross section can be
written as~\cite{Dokshitzer:1991fd}
\begin{equation}
 d\sigma_{Q\to Qg}
 =
 \frac{\alpha_s}{\pi}\,C_F\,
 \frac{\Theta^2\,d\Theta^2}
      {\left(\Theta^2+\Theta_0^2\right)^2}
 \frac{d\omega}{\omega},
 \label{eq:deadcone_cross_section}
\end{equation}
where $\Theta_0$
is the characteristic dead-cone angle defined in Eq.~(\ref{eq:Theta_0}),
$\alpha_s$ is the strong coupling constant, and $C_F=4/3$ is the QCD
colour factor.
For a massless quark, $\Theta_0\to0$,
Eq.~(\ref{eq:deadcone_cross_section}) reduces to the familiar
collinear behaviour
\begin{equation}
 d\sigma_{q\to qg}
 \simeq
 \frac{\alpha_s}{\pi}\,C_F\,
 \frac{d\Theta^2}{\Theta^2}
 \frac{d\omega}{\omega}.
 \label{eq:lightquark_cross_section}
\end{equation}
The radiation probability therefore increases towards the direction
of a massless primary quark. For a massive quark, by contrast,
the gluon radiation is depleted in the immediate vicinity of the heavy-quark direction and
\begin{equation}
d\sigma_{Q\to Qg}
 \longrightarrow0
 \qquad {\rm for}\qquad
 \Theta\longrightarrow0.
\end{equation}
%
%
%
The suppression can be displayed explicitly by dividing the
heavy-quark spectrum by the corresponding massless spectrum. From
Eqs.~(\ref{eq:deadcone_cross_section}) and
(\ref{eq:lightquark_cross_section}) one obtains
\begin{equation}
 {\cal D}(\Theta)
 =
 \frac{d\sigma_{Q\to Qg}}
      {d\sigma_{q\to qg}}
 =
 \left(
 \frac{\Theta^2}
      {\Theta^2+\Theta_0^2}
 \right)^2.
 \label{eq:deadcone_factor}
\end{equation}
Deep inside the dead cone,
$\Theta\ll\Theta_0$,
\begin{equation}
 {\cal D}(\Theta)
 \simeq
 \left(\frac{\Theta}{\Theta_0}\right)^4
 \longrightarrow0,
\end{equation}
while at the characteristic scale
\begin{equation}
 {\cal D}(\Theta_0)=\frac{1}{4}.
 \label{eq:deadcone_at_theta0}
\end{equation}
For $\Theta\gg\Theta_0$, ${\cal D}(\Theta)\to1$.
Thus, $\Theta_0$ specifies the characteristic angular scale of
a continuous suppression which becomes progressively stronger towards
the heavy-quark direction.
The scale $\Theta_0=M_Q/E_Q$ also makes the basic mass and energy
dependence explicit. At fixed energy, the physical angular size of the
dead cone increases with the quark mass, whereas for a fixed quark
flavour it decreases with increasing energy. The comparison of charm,
bottom, and top quarks therefore probes mass-dependent QCD radiation
over a particularly large range of $M_Q$.
Eq.~(\ref{eq:deadcone_cross_section}) describes the elementary
soft-gluon radiation pattern and provides the geometrical basis for
the angular analysis below. The subsequent coherent parton cascade
translates this angular suppression into a modification of the
single-particle momentum distribution. The angular radiation pattern
and the heavy-quark fragmentation function therefore provide
complementary views of the same mass-dependent QCD dynamics.

\subsection{Angular representation in the $(X,Y)$ plane}
\label{subsec:XY}

To compare the radiation patterns of heavy quarks with different
masses and energies, we introduce angular coordinates normalized to
the corresponding dead-cone angle. For a particle emitted at polar
angle $\Theta$ relative to the primary heavy-quark direction and at
azimuthal angle $\phi$, we introduce~\cite{Maltoni:2016ays}
\begin{equation}
 X=
 \frac{\Theta}{\Theta_0}\cos\phi,
 \qquad
 Y=
 \frac{\Theta}{\Theta_0}\sin\phi .
 \label{eq:XY}
\end{equation}
The heavy-quark direction is located at the origin, and the radial
distance is
\begin{equation}
 R\equiv\sqrt{X^2+Y^2}
 =
 \frac{\Theta}{\Theta_0}.
 \label{eq:R}
\end{equation}
Consequently,
\begin{equation}
 X^2+Y^2=1
 \qquad\Longleftrightarrow\qquad
 \Theta=\Theta_0 .
 \label{eq:unitcircle}
\end{equation}

This representation expresses all angular distances in units of the
natural mass-dependent scale $\Theta_0$. The region $R<1$ corresponds
to emission angles below the characteristic dead-cone scale, whereas
$R>1$ corresponds to radiation outside it. A depletion around the
origin in the $(X,Y)$ plane therefore provides a direct geometrical
illustration of the suppression described by
Eq.~(\ref{eq:deadcone_factor}).
We illustrate this representation for the $Z$ pole at
$\sqrt{s}=91.2$~GeV. 
%
%
The Table~\ref{tab:deadcone_angles_LEP} displays
the mass
dependence of the characteristic angular scales. They exhibit the expected hierarchy
\begin{equation}
    \Theta_0^c < \Theta_0^b,
\end{equation}
reflecting the larger bottom-quark mass. In physical angular
coordinates, the characteristic scale is of the order of a few degrees for
charm and about six degrees for bottom at the $Z$ pole.

\begin{table}[H]
\centering
\begin{tabular}{c|c|c|c}
\hline
Quark & $M_Q$ [GeV] & $\Theta_0$ [rad] & $\Theta_0$ [deg] \\
\hline
$c$ & $1.5$ & $3.29\times10^{-2}$ & $1.88^\circ$ \\
$b$ & $4.8$ & $1.05\times10^{-1}$ & $6.03^\circ$ \\
\hline
\end{tabular}

\caption{Characteristic dead-cone angles $\Theta_0=M_Q/E_Q$ for charm and bottom quarks
at the $Z$ pole, using 
$E_Q\simeq45.6$~GeV and 
the kinematic quark masses
employed in the \textsc{Pythia} simulation, $M_c=1.5$~GeV and
$M_b=4.8$~GeV.}
\label{tab:deadcone_angles_LEP}
\end{table}

\begin{figure}[!tb]
    \centering
    \includegraphics[width=\columnwidth]
    {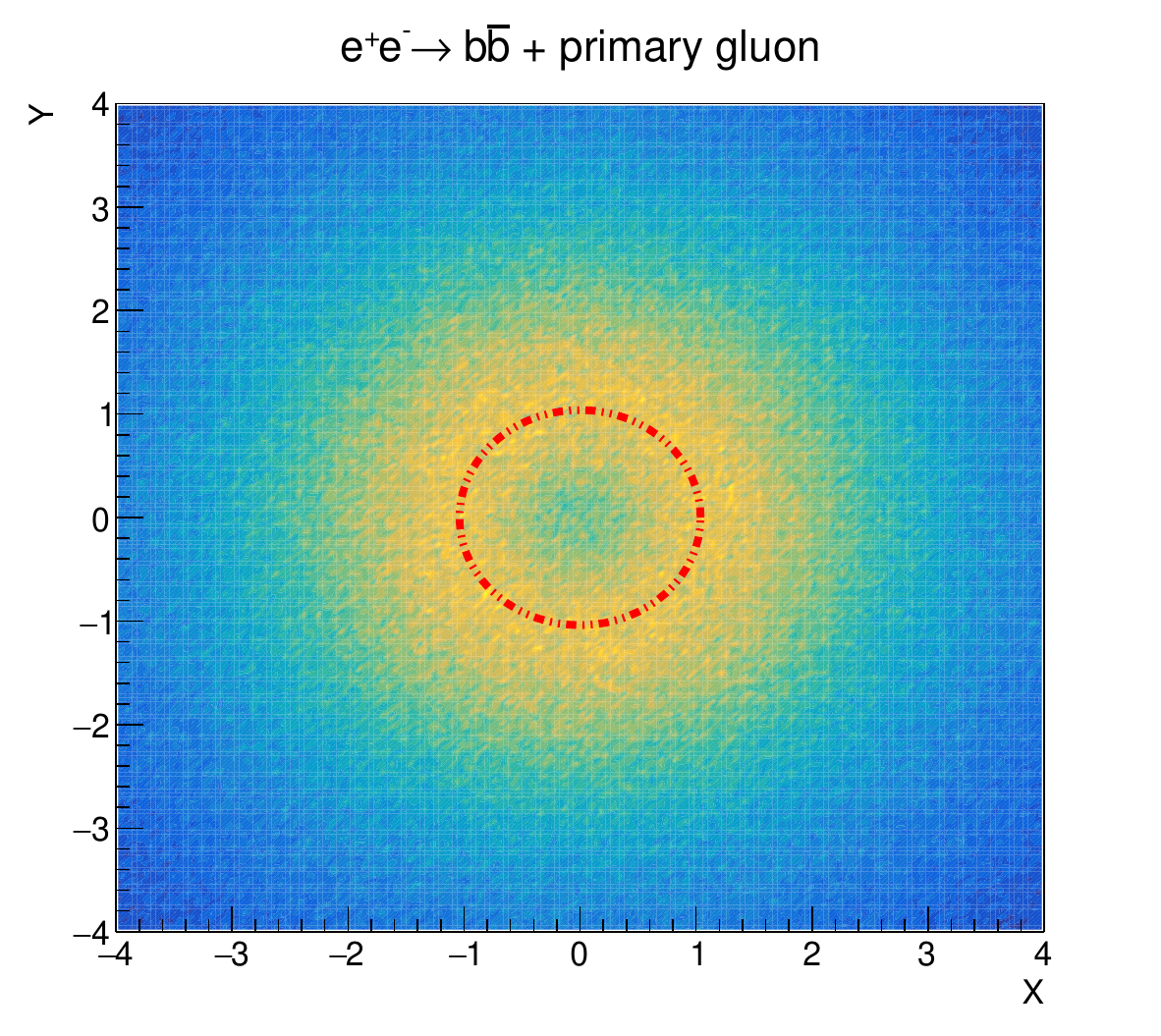}
    \caption{
    Radiation pattern of the primary gluons reconstructed from partons
    in a bottom-quark jet produced in $e^{+}e^{-}$ annihilation at
    $\sqrt{s}=91.2$~GeV, generated with \textsc{Pythia}. The angular coordinates
    $(X,Y)$ are defined in Eq.~(\ref{eq:XY}); the dashed red circle marks
    the dead-cone angle $\Theta_0=M_b/E_b$, see Eq.~(\ref{eq:Theta_0}).
    }
    \label{fig:deadcone_b_LEP}
\end{figure}

For the two-dimensional
illustration in Fig.~\ref{fig:deadcone_b_LEP}, we use the bottom-quark case
as a representative example. Its larger characteristic angle makes the
depleted forward region particularly clear. The angular distribution is
generated at parton level with \textsc{Pythia}~\cite{Bierlich:2022pfr}
and displayed in the normalized $(X,Y)$ coordinates defined in
Eq.~(\ref{eq:XY}).

The angular analysis is performed at parton level in one hemisphere
using the jet reconstruction package
\textsc{FastJet}~\cite{Cacciari:2011ma,Cacciari:2005hq}, with the
$e^+e^-$ generalized-$k_t$ algorithm in inclusive mode and jet radius
$R=0.2$. The $b$ quark jet itself is excluded, and the four-momenta of all
remaining partons in the reconstructed jets are summed to define the
primary-gluon momentum used in the $(X,Y)$ distribution. The reference
circle identifies the characteristic scale $\Theta_0$, but should not
be interpreted as a sharp boundary separating radiating and
non-radiating regions. As follows from the angular dependence in
Eq.~(\ref{eq:deadcone_cross_section}), the suppression becomes strongest
as $\Theta\to0$.

We now turn to the
principal quantitative observable: the single-particle inclusive
momentum distribution.

\subsection{Dead cone in momentum space: MLLA and LEP data}
\label{sec:LEP_MLLA}
The connection between angular space results and momentum space results is particularly
relevant for the experimental analysis.
Through the coherent development of the parton
cascade, the suppression of radiation at small angles modifies the
single-particle inclusive momentum distribution. The heavy-quark mass
therefore leaves a characteristic momentum-space signature in the
form of a depletion of particles carrying large momentum fractions.
The LEP measurements are expressed in terms of the momentum variable
$\xi_p$ at the c.m.s.\ energy $W=91.2$~GeV. The fragmentation function is
defined as
\begin{equation}
    \bar D(\xi,W)
    =
    \frac{1}{2}\frac{1}{\sigma_{\rm tot}}
    \frac{d\sigma^h}{d\xi}(\xi,W),
    \quad
    \bar D(\xi,W)=xD(x,W),
    \label{eq:FF_LEP}
\end{equation}
where we included charged particles only.
The genuine heavy-quark fragmentation functions are reconstructed from
tagged $b\bar b$ and $c\bar c$ events after subtracting the charged
particles originating from decays of the primary heavy hadrons.
The
resulting heavy-quark spectra can then be compared directly with the
light $uds$-quark fragmentation functions measured by DELPHI and
OPAL~\cite{DELPHI:1998cgx,OPAL:1998arz}.


\begin{figure*}[!t]
    \centering
    \begin{tabular}{cc}
        \includegraphics[width=0.47\textwidth]
        {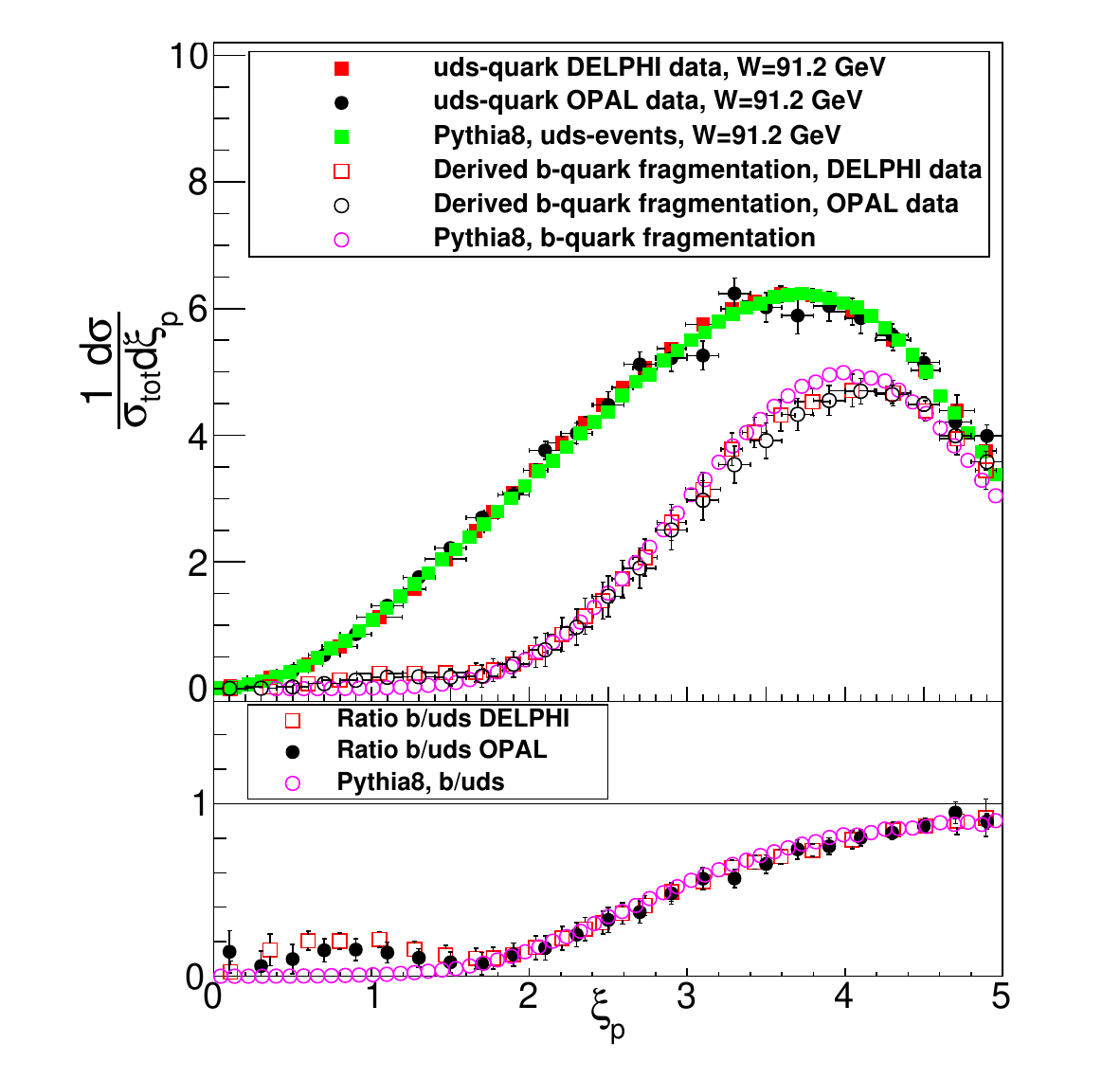}
        &
        \includegraphics[width=0.45\textwidth]
        {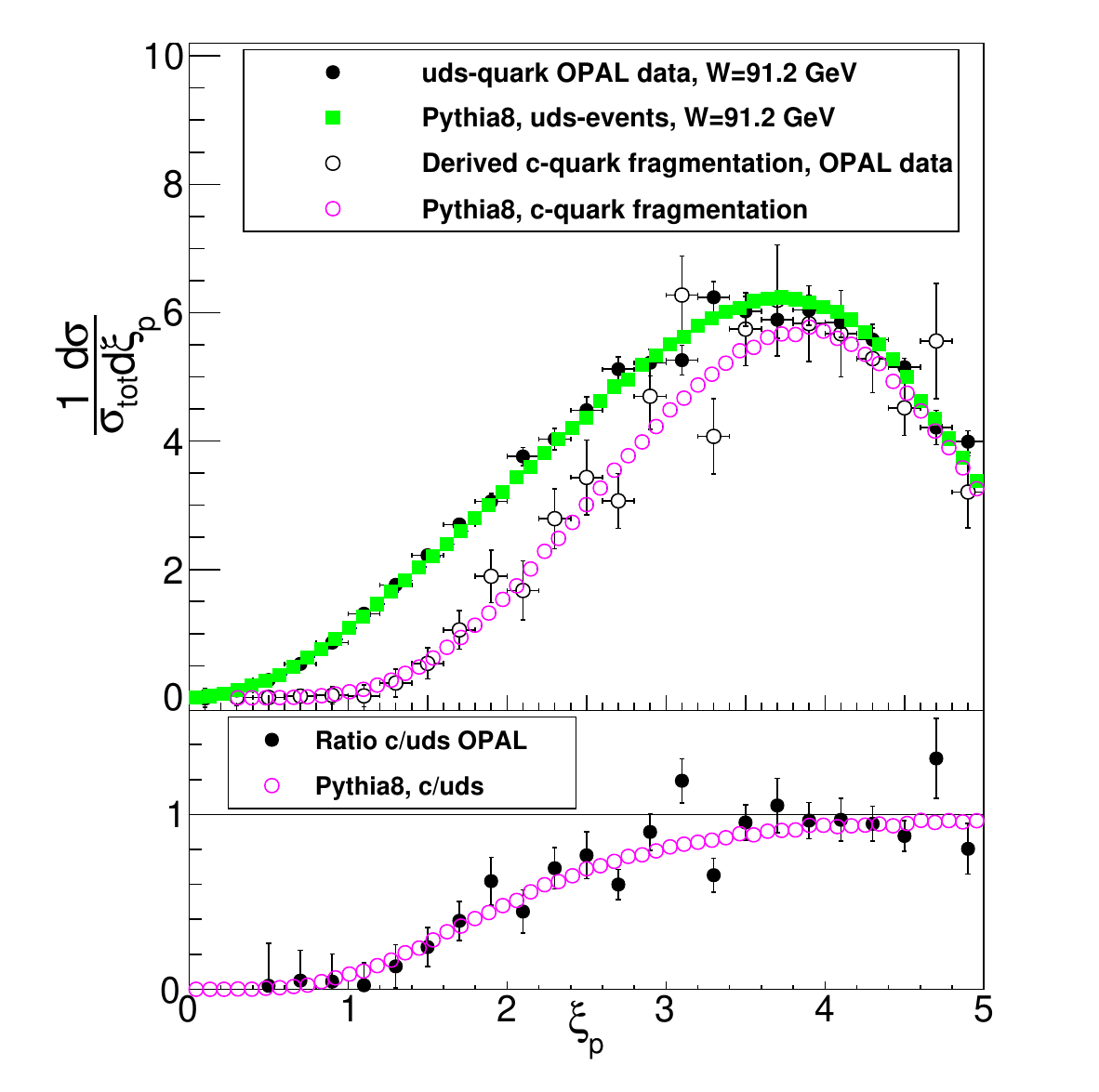}
    \end{tabular}
    \caption{
    Fragmentation functions in $\xi_p=\ln(1/x_p)$ for light
    $uds$ quarks compared with the experimentally reconstructed
    bottom (left) and charm (right) quark fragmentation functions at
    $W=91.2$~GeV. The lower panels show the corresponding ratios of
    heavy- to light-quark fragmentation. A strong depletion develops
    towards small $\xi_p$, i.e.\ large particle momenta, providing
    the momentum-space manifestation of the dead-cone effect.
    }
    \label{fig:exp-deadcone}
\end{figure*}
Fig.~\ref{fig:exp-deadcone} displays the characteristic
momentum-space signature of the dead cone. At large $\xi_p$,
corresponding to soft particles, the heavy- and light-quark spectra
approach one another. Towards smaller $\xi_p$, however, a strong
depletion develops. For bottom quarks the ratio begins to decrease
around $\xi_p\sim5$ and reaches approximately $1/10$ near
$\xi_p\sim2$, whereas for charm the onset of the suppression is
shifted towards smaller $\xi_p$. The effect is established with a
significance larger than $5\sigma$ for both heavy flavours
\cite{Kluth:2023umf}.
A quantitative interpretation can be obtained within the Modified
Leading Logarithmic Approximation
\cite{Dokshitzer:1991fc,Dokshitzer:1991fd}. Within the MLLA estimate
for the inclusive spectrum, the heavy-quark distribution is related
to light-quark spectra through~\cite{Dokshitzer:1991fd}
\begin{equation}
    \bar D_Q(\xi,W)
    =
    \bar D_q(\xi,W)
    -
    \bar D_q\left(
        \xi-\xi_Q,W_0
    \right),
    \label{eq:MLLAeqxi}
\end{equation}
where
\begin{equation}
    W=\sqrt{s},
    \qquad
    W_0=\sqrt{e}M_Q,
    \qquad
    \xi_Q=\ln\frac{1}{\langle x_Q\rangle}.
\end{equation}
The second term in Eq.~(\ref{eq:MLLAeqxi}) represents, within this
MLLA estimate, the radiation which is absent from the heavy-quark jet
as a consequence of the dead cone. Making use of Table~\ref{tab:deadcone_angles_LEP}, for bottom and charm the numerical scales used in the 
analysis are
\begin{equation}
    W_0^{(b)}=8.0~{\rm GeV},
    \qquad
    W_0^{(c)}=2.7~{\rm GeV},
\end{equation}
with
\begin{equation}
    \langle x_b\rangle=0.7092,
    \qquad
    \langle x_c\rangle=0.495,
\end{equation}
corresponding to
\begin{equation}
    \xi_b=0.36,
    \qquad
    \xi_c=0.70,
\end{equation}
using the heavy-quark fragmentation inputs discussed in
Refs.~\cite{Dokshitzer:2005ri,DELPHI:2011aa,Baines:2006uw}.
\begin{figure*}[!t]
    \centering
    \begin{tabular}{cc}
        \includegraphics[width=0.47\textwidth]
        {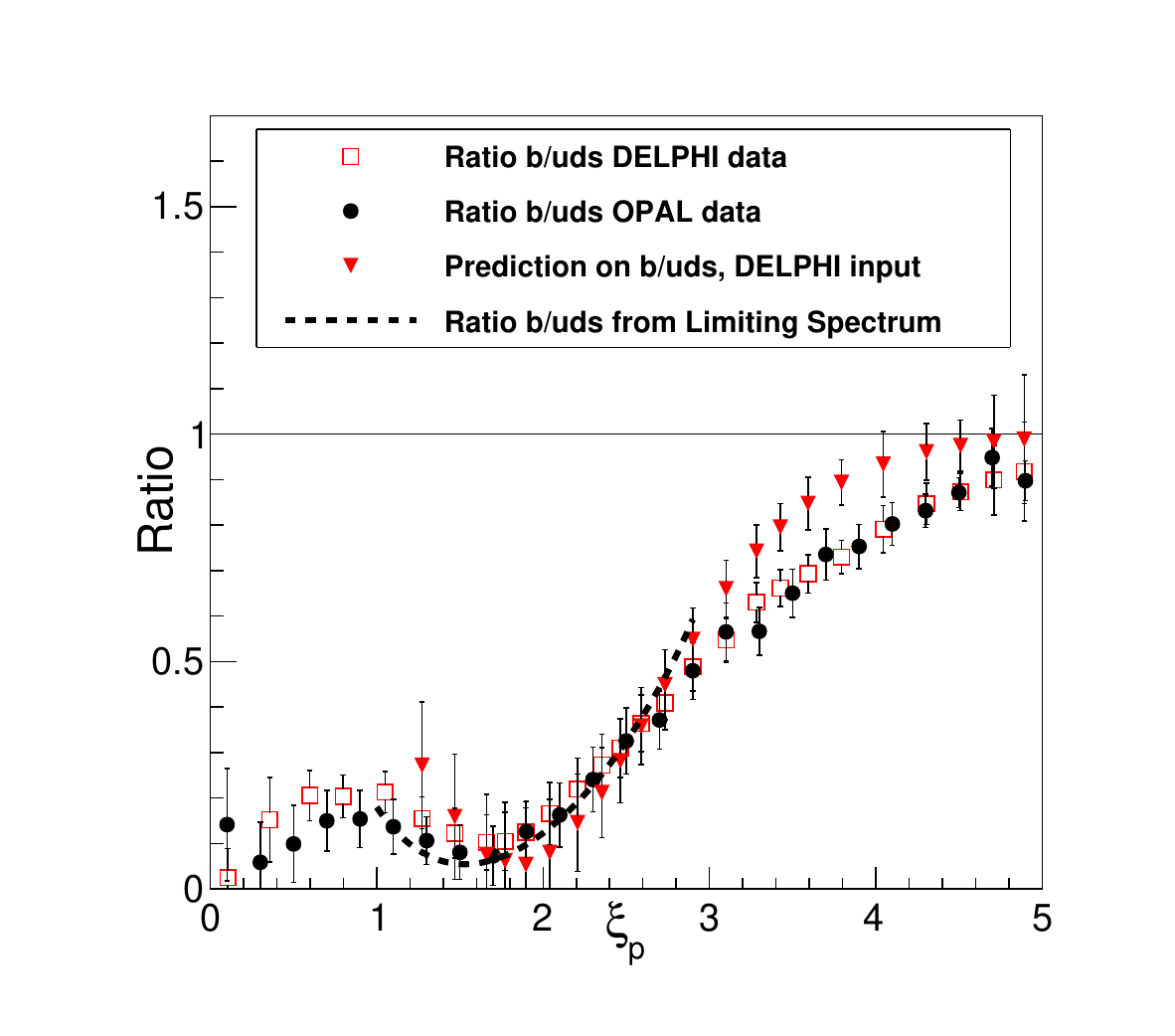}
        &
        \includegraphics[width=0.47\textwidth]
        {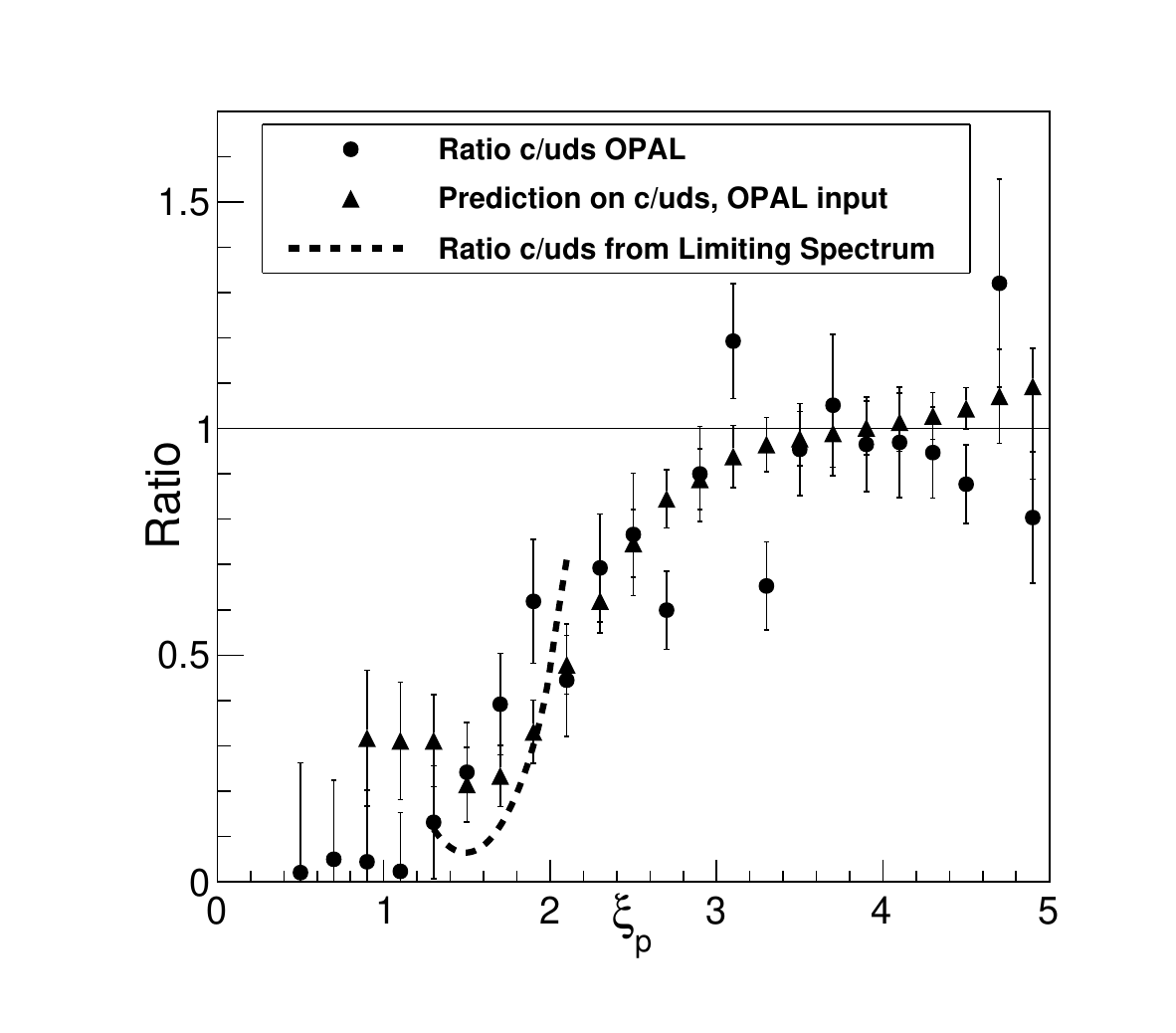}
    \end{tabular}
    \caption{
    Ratios of the bottom (left) and charm (right) fragmentation
    functions to the light $uds$-quark fragmentation function at
    $W=91.2$~GeV, compared with the MLLA expectations obtained using
    experimental input and the Limiting Spectrum.
    }
    \label{fig:exp-deadcone_ratios}
\end{figure*}
As shown in Fig.~\ref{fig:exp-deadcone_ratios}, the MLLA estimate
reproduces the observed suppression over the central kinematic
region, approximately $1\lesssim\xi_p\lesssim3$ for bottom and
$1\lesssim\xi_p\lesssim2$ for charm. The different suppression
profiles are naturally connected with the different mass scales
$W_0=\sqrt{e}M_Q$ and hence with the low-energy light-quark spectra
entering Eq.~(\ref{eq:MLLAeqxi}). Finally, towards smaller $\xi_p$, the charm- and bottom-quark spectra are strongly depleted relative to the light $uds$-quark spectrum. For
bottom quarks, the ratio indicates a suppression by an order of magnitude
\begin{equation}
    \frac{\bar D_b}{\bar D_{uds}}
    \sim 10^{-1}
\end{equation}
around $\xi_p\simeq2$. 
For charm quarks the suppression sets in at
smaller $\xi_p$ and becomes increasingly strong towards the
high-momentum kinematic boundary.

The comparison should not be extended indiscriminately to the
kinematic boundaries. Equation~(\ref{eq:MLLAeqxi}) is an approximate
MLLA relation for the inclusive differential spectrum, and deviations
are expected where the approximations underlying the perturbative
description cease to be reliable.
The LEP results thus establish a direct connection between the two
manifestations of the dead-cone phenomenon. 
The
suppression can be visualized geometrically through the depletion of
radiation at small angles. After the development of the coherent QCD
cascade, the same mass effect appears in momentum space as a strong
suppression of high-momentum particles.
Having established this correspondence for charm and bottom at the
$Z$ pole, we next extend the analysis, based on the \textsc{Pythia} Monte
Carlo simulation, to a common energy of $\sqrt{s}=1$~TeV, including the
top quark.

\section{Extension to the top-quark mass scale}
\label{sec:top}

We now consider
\begin{equation}
    e^+e^-\to Q\bar Q,
    \qquad Q=c,b,t,
\end{equation}
at $\sqrt{s}=1$~TeV. This comparison covers more than two orders of
magnitude in heavy-quark mass. The characteristic angular scale
remains that of Eq.~(\ref{eq:Theta_0}),
$\Theta_0=M_Q/E_Q$, so that at fixed energy the physical dead-cone
angle increases strongly from charm to bottom and to top.
The top-quark case introduces an essential complication. With a mass
of about $m_t\simeq172.5$~GeV, the top quark is the heaviest known
elementary particle and has an exceptionally large weak-decay rate.
Its dominant decay mode is
\begin{equation}
    t\to bW ,
\end{equation}
as a consequence of its large mass and the CKM hierarchy
$|V_{tb}|\gg |V_{ts}|,|V_{td}|$. The measured top-quark decay width is~\cite{ParticleDataGroup:2024cfk}
\begin{equation}
    \Gamma_t = 1.42^{+0.19}_{-0.15}~{\rm GeV},
\end{equation}
corresponding through $\tau_t=\hbar/\Gamma_t$ to the characteristic
lifetime
\begin{equation}
    \tau_t \simeq 4.6\times10^{-25}~{\rm s}.
\end{equation}
This lifetime is considerably shorter than the characteristic
hadronization time scale, $\tau_{\rm had}\sim1/\Lambda_{\rm QCD}
\sim10^{-24}$--$10^{-23}$~s. The top quark therefore decays before it
can hadronize into an ordinary top-flavoured hadron. This property
distinguishes it from charm and bottom quarks and makes its radiation
pattern directly sensitive to the interplay between production and
decay.

The short top lifetime also prevents the formation of the narrow,
long-lived quarkonium spectrum familiar from charmonium and
bottomonium. 
\subsection{Heavy-quark mass hierarchy and the stable top at $\sqrt{s}=1$ TeV}

\begin{figure}[!tb]
    \centering
    \includegraphics[width=\columnwidth]
    {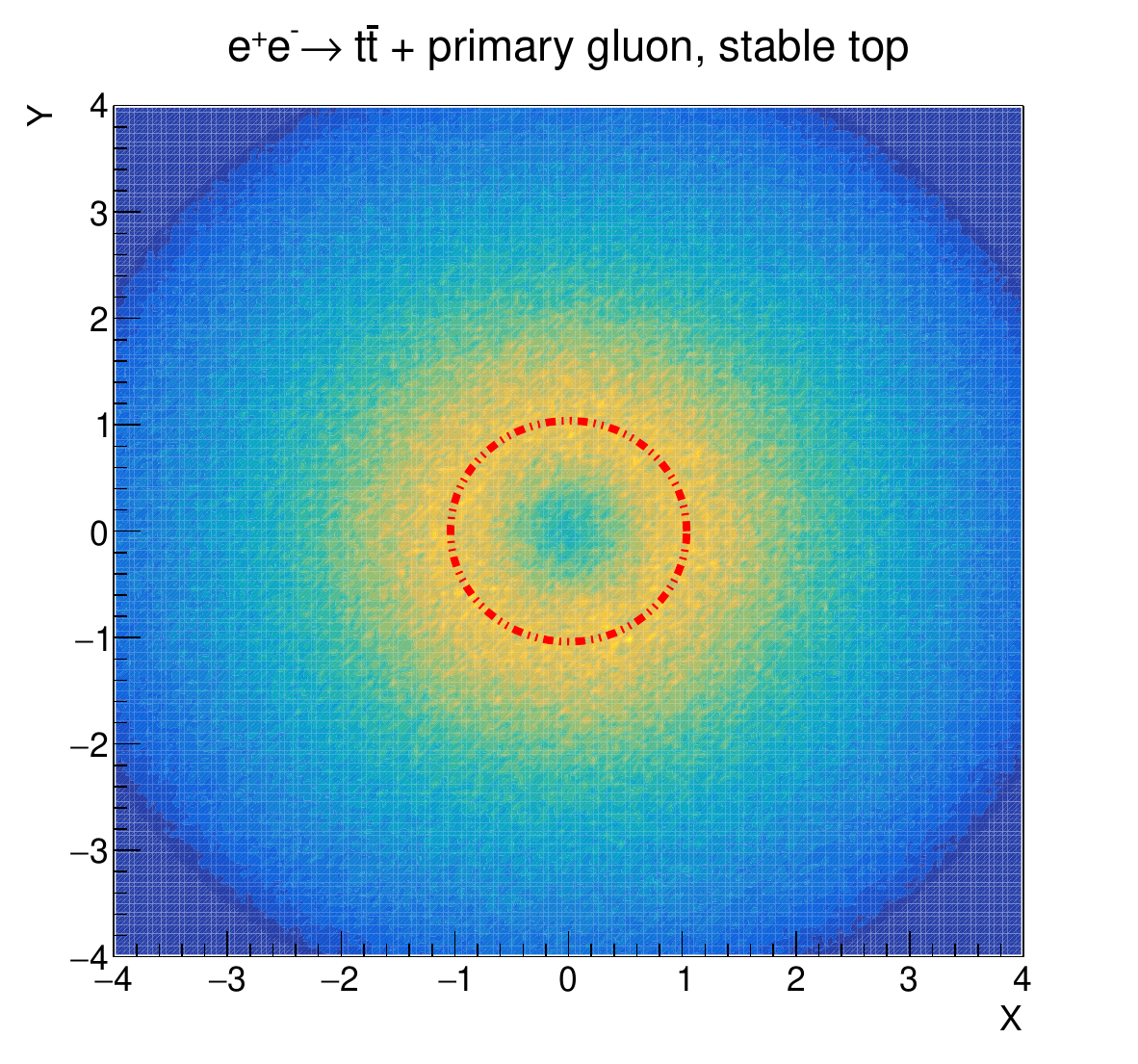}
    \caption{
    Radiation pattern of the primary gluons reconstructed from partons
    in the rescaled angles $X$ and $Y$ as in Fig.~\ref{fig:deadcone_b_LEP},
    but for a hypothetical stable top quark in $e^{+}e^{-}$ annihilation
    at $\sqrt{s}=1$~TeV. The stable-top sample isolates the primary
    top-quark radiation pattern before the effects associated with the
    physical top decay are introduced.
    }
    \label{fig:deadcone_stable_top_1TeV}
\end{figure}
\label{subsec:cbt_1TeV}

We begin by considering the heavy-quark mass hierarchy at
$\sqrt{s}=1$~TeV. Since no experimental data are presently available
for such a study at this energy, the results presented here are
predictions based on Monte Carlo simulations.
At the common quark energy $E_Q\simeq\sqrt{s}/2=500$~GeV, the
characteristic dead-cone angle $ \Theta_0=\frac{M_Q}{E_Q}$
%
%
increases strongly with the heavy-quark mass,
see Table~\ref{tab:deadcone_angles}. 
They illustrate directly the hierarchy
\begin{equation}
    \Theta_0^c \ll \Theta_0^b \ll \Theta_0^t,
\end{equation}
which reflects the underlying mass hierarchy
$m_c\ll m_b\ll m_t$.

\begin{table}[!tb]
\centering
\begin{tabular}{c|c|c|c}
\hline
Quark & $M_Q$ [GeV] & $\Theta_0$ [rad] & $\Theta_0$ [deg] \\
\hline
$c$ & $1.5$ & $3.00\times10^{-3}$ & $0.172^\circ$ \\
$b$ & $4.8$ & $9.60\times10^{-3}$ & $0.550^\circ$ \\
$t$ & $172.5$ & $3.45\times10^{-1}$ & $19.8^\circ$ \\
\hline
\end{tabular}

\caption{Characteristic dead-cone angles for charm, bottom, and top
quarks at $E_Q\simeq500$~GeV. The angles are calculated with the
kinematic quark masses used in the \textsc{Pythia} simulation,
$M_c=1.5$~GeV, $M_b=4.8$~GeV, and $M_t=172.5$~GeV.}
\label{tab:deadcone_angles}
\end{table}


Next we concentrate on the case of the top quark, which, with the
largest mass, introduces the additional complication of the top-quark
decay.
As a reference, we first consider a hypothetical stable top quark.
At parton level, the final state is reconstructed with \textsc{FastJet}
using the $e^+e^-$ generalized-$k_t$ algorithm with $R=0.2$. The top quark jet
itself is excluded, and the four-momenta of all remaining partons in the
reconstructed jets are summed to define the primary-gluon momentum.
Figure~\ref{fig:deadcone_stable_top_1TeV} displays the corresponding
radiation pattern in the normalized $(X,Y)$ plane. Since the angular
variables are rescaled by the characteristic dead-cone angle
$\Theta_0$, this distribution can be directly compared with the
bottom-quark distribution in Fig.~\ref{fig:deadcone_b_LEP}. Despite the
very different physical angular scales, the two distributions exhibit
a similar structure in the normalized variables. The depletion around
the top-quark direction provides the clean reference configuration
against which the physical unstable-top case will be compared.

\subsection{Unstable top quark: the physical decay}
\label{subsec:unstable_top}

\begin{figure}[!tb]
    \centering
    \includegraphics[width=1.0\columnwidth]
    {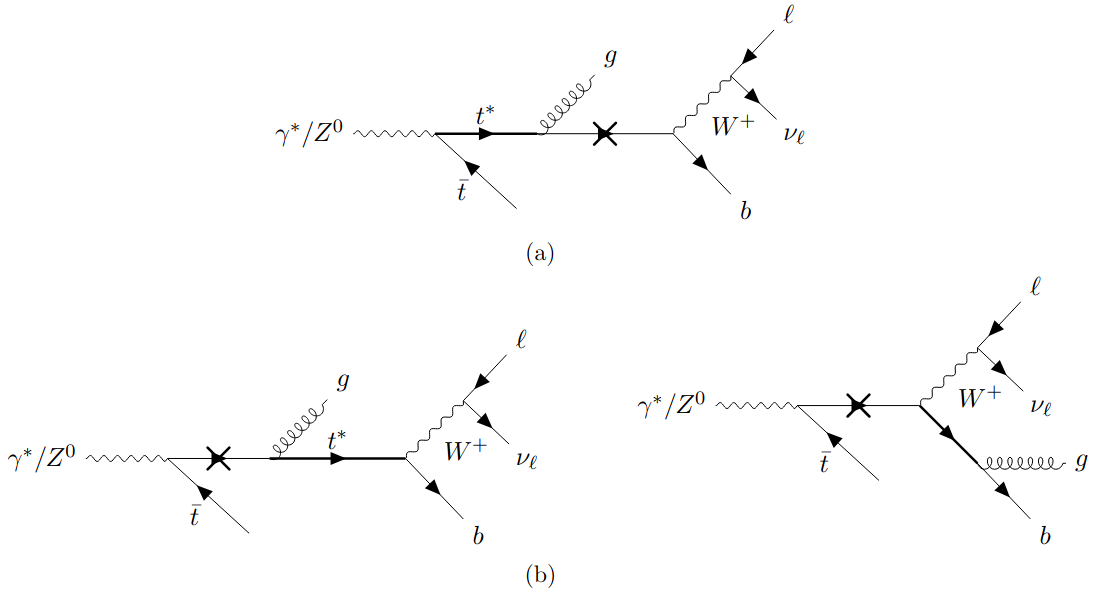}
    \caption{
    Feynman diagrams for gluon emission in
    $e^+e^-\to t\bar t\to bW^+\bar bW^-$ relevant to the
    top-quark hemisphere. Panel (a) represents gluon radiation
    associated with top-quark production (dead-cone signal), while panel (b) represents
    radiation associated with the decay $t\to bW$, including emission
    from the daughter $b$ quark (background). Heavy lines denote off-shell quarks.
    The decomposition follows the leading-order treatment of
    Ref.~\cite{Orr:1992uv}.
    }
    \label{fig:Feynman}
\end{figure}

The Standard Model predicts an unstable top quark which decays before
hadronization, predominantly through $t\to bW$. We consider, for
simplicity, the leptonic channel
\begin{equation}
    t\to bW^+\to b\ell^+\nu_\ell ,
    \label{eq:top_leptonic_decay}
\end{equation}
together with the charge-conjugate decay of the antitop. The essential
complication relative to the stable-top reference is the coloured
daughter $b$ quark and the associated coherent decay radiation.
At leading order in $\alpha_s$, the origin of the relevant radiation
components can be understood from
\begin{equation}
 e^+e^-\to t\bar t\to bW^+\,\bar bW^-+g .
\end{equation}
The corresponding diagrams are shown in Fig.~\ref{fig:Feynman}.
%
%
For soft-gluon emission the corresponding distribution can be written
in the form~\cite{Orr:1992uv,Jikia:1991yd}
\begin{equation}
 \frac{1}{\sigma_0}
 \frac{d\sigma_g}
 {d\omega\,d\cos\Theta_g\,d\phi_g}
 =
 \frac{\alpha_s C_F}{4\pi^2}\,
 \omega\,F ,
 \label{eq:unstable_top_gluon}
\end{equation}
where, retaining the components relevant to the present discussion,
\begin{equation}
 F=|A|^2+|B_1|^2+\cdots .
 \label{eq:AB_decomposition}
\end{equation}
As will be detailed below, the amplitude $A$ describes the radiation from the
$\widehat{t\bar t}$ colour dipole and contains the primary top-quark
dead cone. The amplitude $B_1$ describes radiation associated with
the $\widehat{tb}$ dipole generated by the top decay. The corresponding
$\widehat{\bar t\bar b}$ contribution belongs to the opposite
hemisphere.
It is therefore not sufficient to regard the $b$ jet as an
independent secondary jet which can simply be removed. The coherent
$\widehat{tb}$ dipole itself radiates into the top-quark hemisphere.

The interference terms left out in \eqref{eq:AB_decomposition} correspond 
to contributions, which depend on the top-width $\Gamma_t$; these 
contributions are found to be negligible (typically $\lesssim 10\%$) 
for the present analysis~\cite{Kluth:2025xpj}.

\subsubsection{Angular structure of the decay radiation}
\label{subsubsec:tb_dipole}

In the high-energy and small-angle limit, the primary-top component
contains the familiar dead-cone factor,
\begin{equation}
 \omega^2 |A|^2
 \simeq
 \frac{4\Theta_g^2}
      {(\Theta_g^2+\Theta_0^2)^2},
 \qquad
 \Theta_0\simeq\frac{m_t}{E_t},
 \label{eq:A_top}
\end{equation}
where $\Theta_g$ is measured relative to the top-quark direction. This term is the one appearing in the differential cross section in Eq.~(\ref{eq:deadcone_cross_section}) as expected.
Let $\Theta_b$ denote the angle between the $b$-quark and top-quark
directions. After integration over the gluon angles, the leading
dependence of the decay contribution may be represented schematically
as
\begin{equation}
 \omega^2\langle |B_1|^2\rangle
 \propto
 1-\beta_t\beta_b\cos\Theta_b ,
 \label{eq:B1_angle}
\end{equation}
with heavy quark velocities $\beta_Q$, up to mass-suppressed terms and a smooth angular factor. At high
energies this contribution is strongly reduced when the $b$ quark
approaches the top direction; in the corresponding limiting form,
\begin{equation}
 \omega^2\langle |B_1|^2\rangle
 \propto 1-\cos\Theta_b .
 \label{eq:B1_forward}
\end{equation}
This behaviour provides the basis for the reconstruction of the momentum spectrum in the top quark fragmentation.
Although the decay radiation cannot in general be removed on an
event-by-event basis, its contribution can be studied at finite
$b$-quark decay angles $\Theta_b$ and extrapolated towards the forward limit
$\Theta_b\to0$, where the $\widehat{tb}$ contribution is strongly
suppressed.
The subsequent parton cascade preserves an important trace of this
dipole structure through angular ordering.
Radiation initiated by the $b$ quark is preferentially confined to
the angular region on the $b$ side of the event. We therefore rotate
the event such that the $b$-jet direction lies along the negative
$X$ axis. Consequently, in the region $X>0$, opposite to the $b$ jet, the background coming from the b-initiated cascade is suppressed.

\subsubsection{Separating production and decay radiation}
\label{subsubsec:top_decay_separation}

A complementary separation of the two radiation mechanisms is
obtained from the reconstructed invariant masses~\cite{Maltoni:2016ays}. 
For radiation predominantly associated
 with the production stage, $t^*\to tg$, the invariant
mass $M(b\ell\nu)$ remains close to $m_t$. For radiation that occurs in
decay $t\to b\ell\nu+g$, the combination $M(b\ell\nu+g)$ instead
reconstructs the top mass.

The unstable-top analysis is performed within the same FastJet-based
jet-reconstruction framework introduced above, using the
$e^+e^-$ generalized-$k_t$ algorithm with $R=0.2$. This reconstruction
is used throughout the analysis, in particular to identify the gluon
jets entering the invariant-mass selections and the characteristic
bands in the $(X,Y)$ plane discussed below. It also provides a basis
for the subsequent angular distributions of the reconstructed gluon
jets.


\begin{figure}[!tb]
    \centering
    \includegraphics[width=\columnwidth]
    {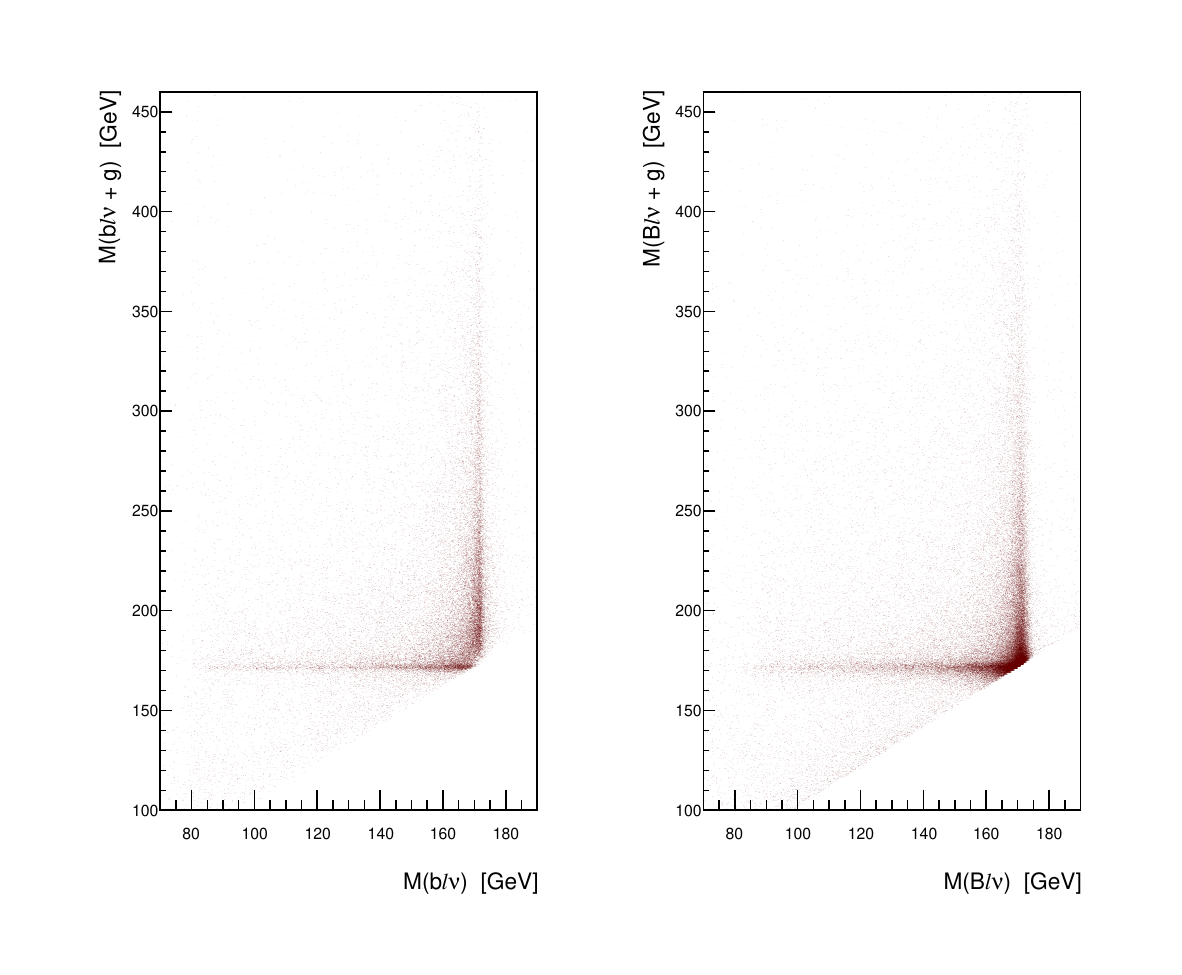}
    \caption{
Correlation of the reconstructed invariant masses
$M(be\nu_e+g)$ and $M(be\nu_e)$ in
$e^+e^-\to t\bar t+X$, with $t\to be\nu_e$, at
$\sqrt{s}=1$~TeV. The left panel corresponds to partons and the right
panel to hadrons, with the $b$ quark replaced by the corresponding
$B$ hadron. The vertical structure near
$M(be\nu_e)\simeq m_t$ is associated predominantly with
production-stage radiation $t^*\to tg$, whereas the horizontal
structure near $M(be\nu_e+g)\simeq m_t$ enhances radiation from the
top decay.
}
    \label{fig:top_stripes}
\end{figure}
In Fig.~\ref{fig:top_stripes} we show the scatter plots for these two masses for partons (left panel) and hadrons (right panel).
In both cases, the two bands provide a useful
phenomenological separation of the two components. The separation is
not exact, since the complete parton shower contains multiple
emissions and interference effects, but it provides a transparent
connection between the leading-order dipole picture and the full
\textsc{Pythia} event. The vertical stripe includes the genuine dead-cone signal associated
with primary top-quark radiation, whereas the horizontal stripe
corresponds to decay radiation that contaminates and partially fills
the dead-cone region. The signal is therefore selected by requiring the reconstructed
invariant mass to lie in a narrow window around the top-quark mass~\cite{Maltoni:2016ays,Kluth:2025xpj},
\[
165 \leq M(b\ell\nu) \leq 175~\mathrm{GeV}.
\]
for partons and hadrons.

\begin{figure*}[!t]
    \centering
    \includegraphics[width=\textwidth]{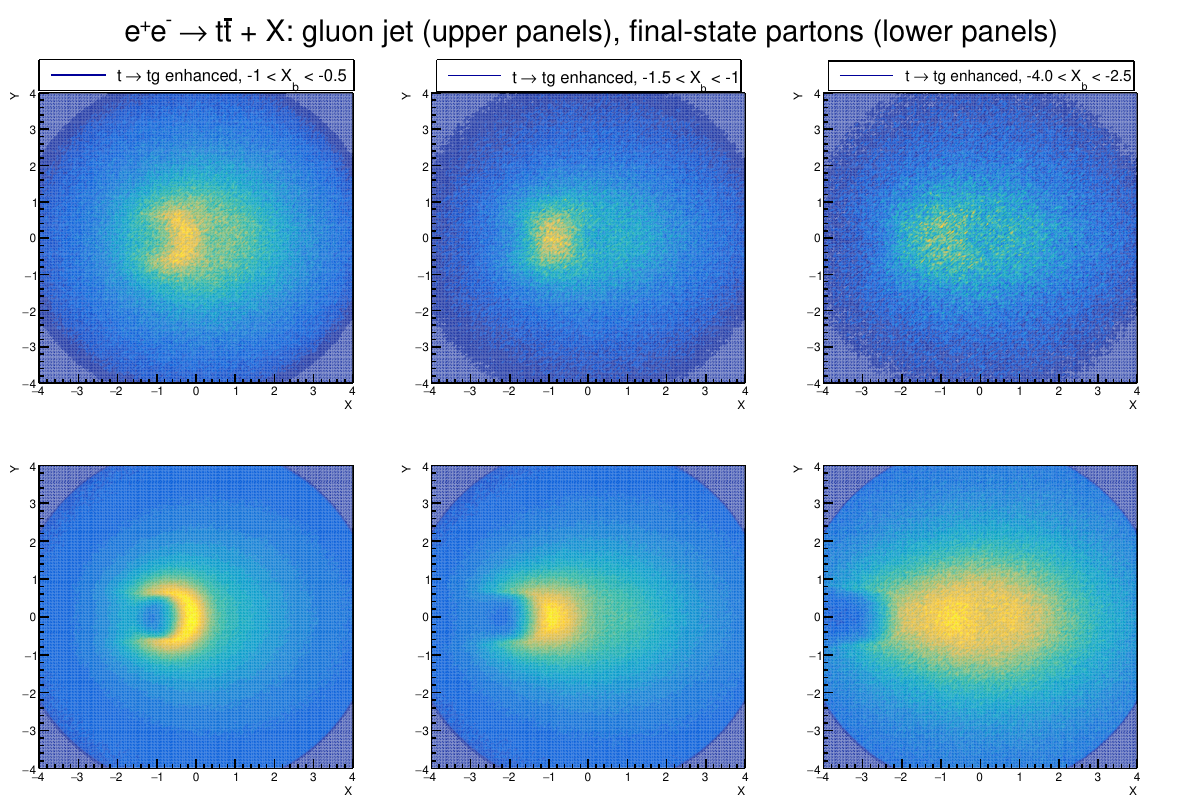}
    \caption{
    Angular distributions in the $(X,Y)$ plane for an unstable top quark
    at $\sqrt{s}=1$~TeV after the top-mass selection
    $165\leq M(b\ell\nu)\leq175$~GeV: in the upper panels for the
    reconstructed gluon jets, in the lower panels for the corresponding
    final-state partons emerging from the jet evolution. 
    The three columns correspond to different b-quark angles $X_b=\Theta_b/\Theta_0$, from left to right $-1<X_b<-0.5$, $-1.5<X_b<-1$, and
    $-4.0<X_b<-2.5$, respectively. The $b$-quark direction is oriented
    along the negative $X$-axis, b-quark jets are removed (dark circular areas on the left sides), gluon jets selected with $E_g>10$ GeV.
    }
    \label{fig:scatter_gluon}
\end{figure*}

The corresponding angular distributions are displayed in
Fig.~\ref{fig:scatter_gluon}, with the reconstructed gluon jets in the
upper panels and the partons emerging from the jet evolution in the
lower panels. The three columns correspond to different intervals of the
$b$-quark coordinate $X_b$, which controls the relative orientation of the
bottom- and top-quark directions.
%
%
The upper panels of Fig.~\ref{fig:scatter_gluon} retain the broad radiation
structure of the reconstructed gluon jets around the $\widehat{t\bar t}$
dipole, but the central depletion at $X=Y=R=0$ is much reduced in
comparison to the stable top quark in Fig.~\ref{fig:deadcone_stable_top_1TeV}.
The lower panels show the corresponding distributions for final-state
partons; here, the central suppression has almost disappeared. In the
lowest-$|X_b|$ interval, the strong radiation from the $b$ quark, visible as the
yellow crescent, enters the right-hand side. This additional component is
associated with the coherent radiation generated by the decay of the top-quark
$t\to bW$.

\begin{figure*}[!t]
    \centering
    \begin{tabular}{cc}
        \includegraphics[width=0.47\textwidth]{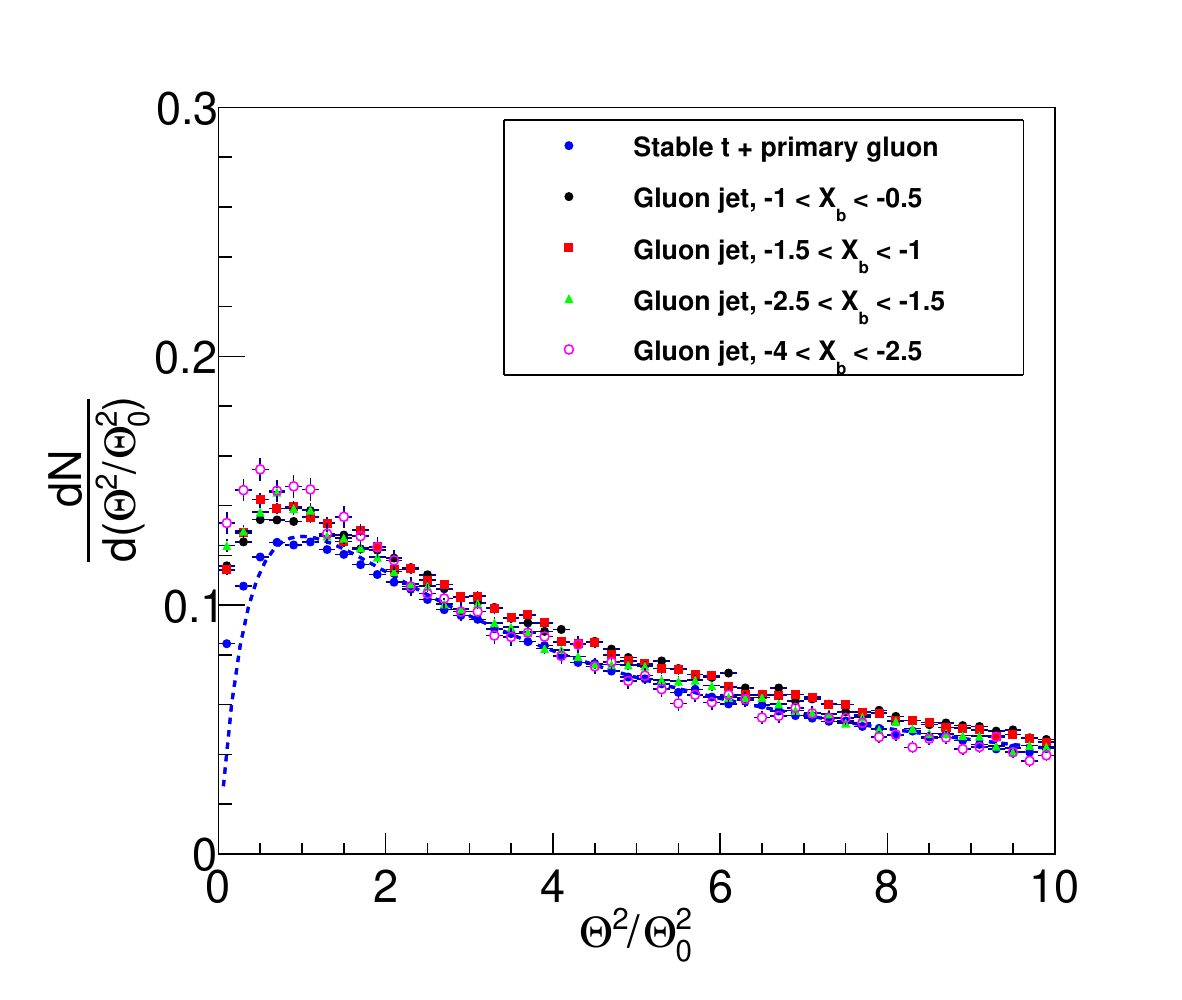}
        &
        \includegraphics[width=0.47\textwidth]{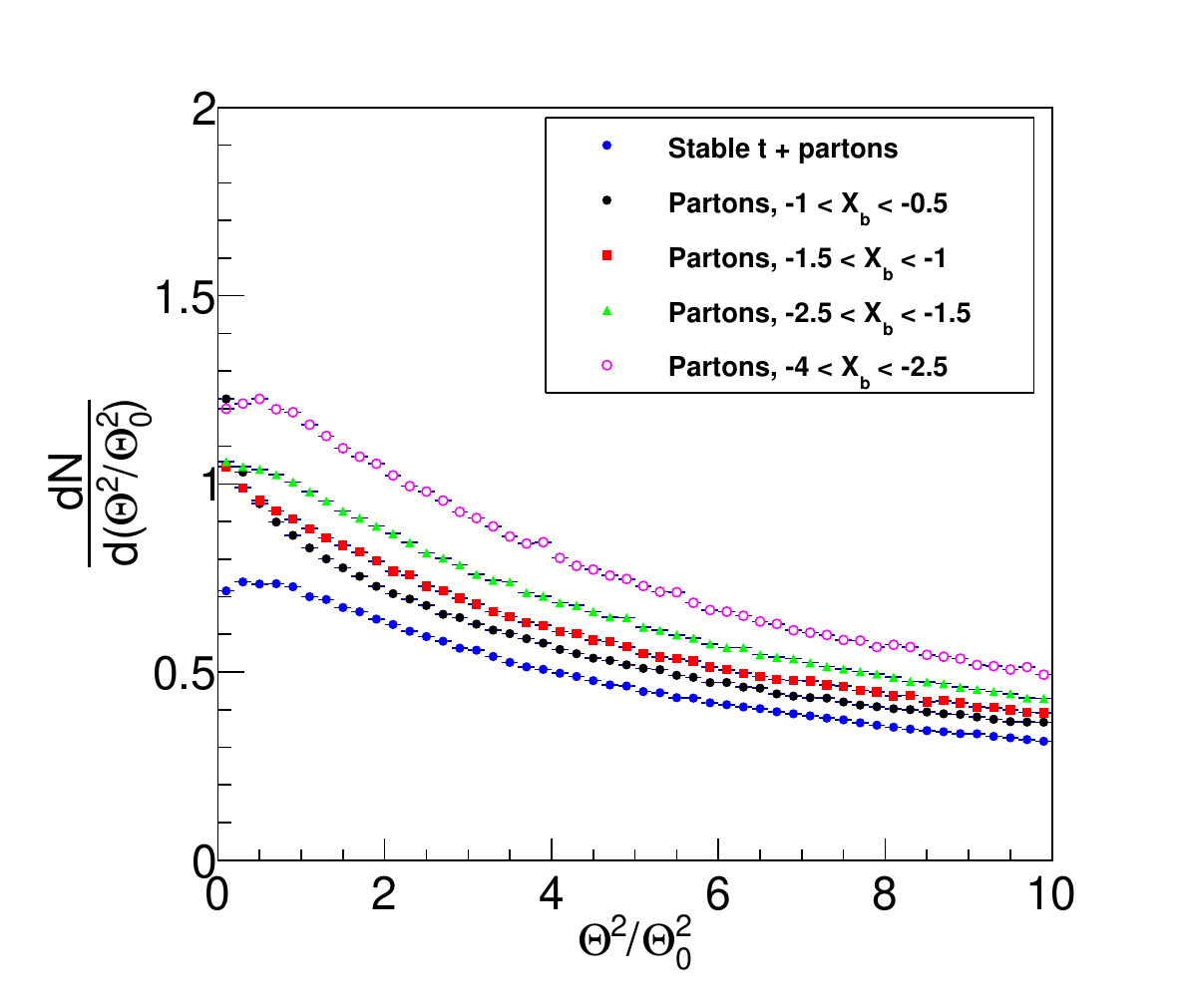}
    \end{tabular}
    \caption{ 
    Distributions in angular variable $\Theta^{2}/\Theta_{0}^{2}=X^2+Y^2$, a one-dimensional projection of 
    Fig.~\ref{fig:scatter_gluon}: in the left panel for the reconstructed-gluon-jets in different b-quark angle $X_b$ intervals, as indicated, corresponding to
    the upper row of Fig.~\ref{fig:scatter_gluon}, 
    in the right panel for 
    the corresponding final-state-parton distributions.
    The stable-top distributions provide the references for
    the primary top-quark radiation. The curve in the left panel represents the gluon distribution in Eq.~\eqref{eq:deadcone_cross_section}. 
    }
    \label{fig:ang_dist_combined}
\end{figure*}

A corresponding one-dimensional view of the radiation pattern of these
panels is shown in Fig.~\ref{fig:ang_dist_combined} as a function of the
normalized squared angle $X^2+Y^2=\Theta^{2}/\Theta_{0}^{2}$. The left panel shows
the radial projections of the reconstructed-gluon-jet distributions,
including the three intervals shown in the upper row of
Fig.~\ref{fig:scatter_gluon} and the additional intermediate interval
$-2.5<X_b<-1.5$. 
The data indicate a small dip at $\Theta^2/\Theta_0^2=0$, but less pronounced
than in the case of the stable top quark. The curve represents the
lowest-order result of Eq.~(\ref{eq:deadcone_cross_section}). In this
analysis only gluon jets with energy $E_g>10$~GeV are included,
following Ref.~\cite{Kluth:2025xpj}.

The right panel shows the corresponding
final-state-parton distributions. At higher values of
$\Theta^{2}/\Theta_{0}^{2}$, the unstable-top distributions evolve
smoothly towards the stable-top reference as the $b$-quark decay angle
$X_b=\Theta_b/\Theta_0$ decreases. However, at small angles $\Theta^{2}/\Theta_{0}^{2}$, 
no corresponding convergence towards the stable-top distribution
is observed, rather a divergence; in the parton-level distributions, the strong radiation from the
$b$-initiated cascade at $X_b<0$ progressively enters the depleted region. Thus, an
extrapolation towards the primary top-quark radiation cannot be performed
reliably from the angular distributions at finite $X_b$ in this region. This motivates
the complementary study of the momentum distributions, for which 
such an extrapolation turns out to give reasonable results.

\subsection{Reconstruction of the primary parton momentum spectrum in the top-quark jet}
\label{subsec:top_momentum_reconstruction}

\begin{figure}[!tb]
    \centering
    \includegraphics[width=1.09\columnwidth]
    {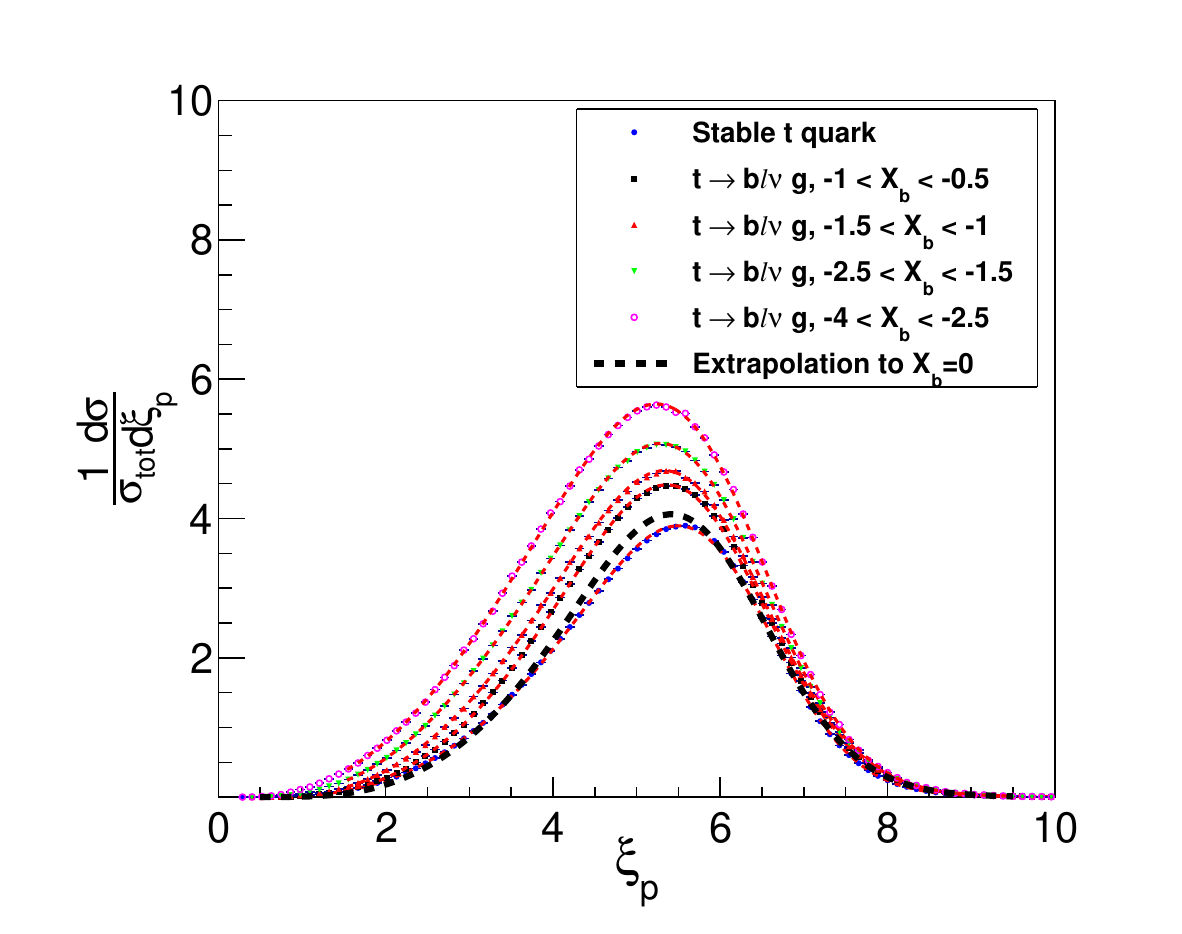}
    \caption{
    Parton momentum distributions in
    $\xi_p=\ln(1/x_p)$ for physical top-quark decays at
    $\sqrt{s}=1$~TeV, determined for different intervals of the
    $b$-jet coordinate $X_b$. Only partons in the half-plane
    $X_p>0$ opposite to the $b$ jet are used, and the distributions
    are multiplied by two. The spectra are fitted with a
    distorted-Gaussian parametrization and extrapolated to $X_b=0$.
    The extrapolated distribution is compared with the independently
    generated stable-top reference.
    }
    \label{fig:multiple_xi_partons}
\end{figure}

The angular analysis suggests a practical way to recover the momentum
distribution associated with the primary top quark. The event is
oriented with the $b$ jet along the negative $X$-axis, and only partons in the
opposite half-plane,
\begin{equation}
    X_p>0,
\end{equation}
are used, where $X_p$ denotes the horizontal angular coordinate of a
parton in the rotated $(X,Y)$ plane. The resulting distribution is
multiplied by two to represent the complete top-quark hemisphere.
%
%
%
We denote by $X_b$ the signed, rescaled horizontal coordinate of the
$b$ jet in the same rotated angular plane. The forward limit
$\Theta_b\to0$ therefore corresponds to $X_b\to0$. The momentum spectra
of partons are determined in several intervals of $X_b$ and extrapolated
to the limit $X_b=0$, where the $\widehat{tb}$ contribution is expected to be
strongly reduced, see Fig.~\ref{fig:multiple_xi_partons}.


\begin{figure}[!tb]
    \centering
    \includegraphics[width=0.82\columnwidth]
    {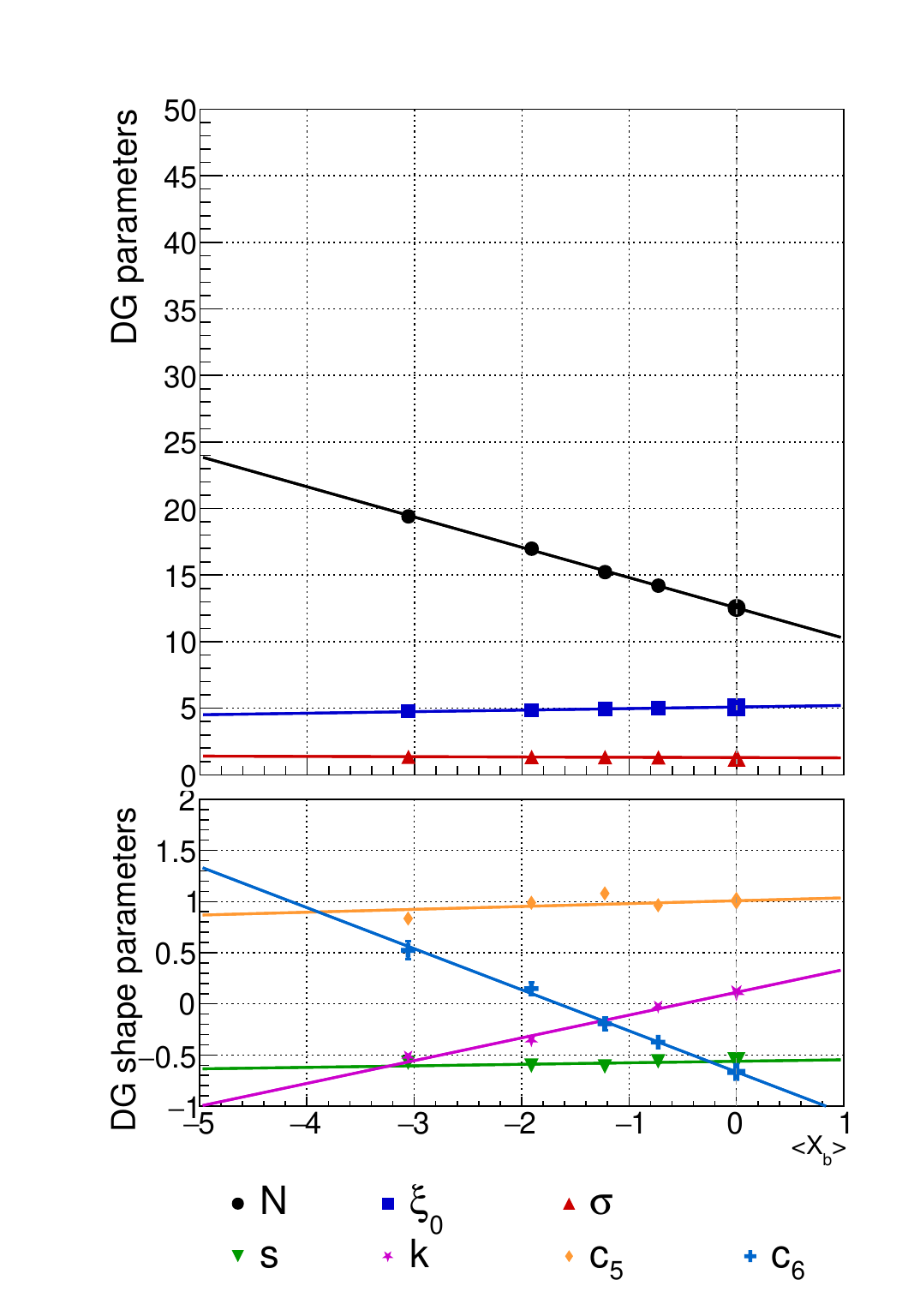}
    \caption{
    Dependence on the $b$-jet coordinate $X_b$ of the parameters
    obtained from the distorted-Gaussian fits to the parton momentum
    distributions shown in Fig.~\ref{fig:multiple_xi_partons}. The
    fitted parameter values are extrapolated to $X_b=0$, corresponding
    to the forward $b$-quark configuration.
    }
    \label{fig:xi_fit_parameters}
\end{figure}
The dominant change of the $\xi_p$ distributions in
Fig.~\ref{fig:multiple_xi_partons} is in their normalization, whereas
their shapes vary more weakly. This behaviour is qualitatively
consistent with the leading-order result in Eq.~(\ref{eq:B1_forward}), in which the gluon-energy
and angular dependences approximately factorize.
The $\xi_p$ distributions are described by a distorted-Gaussian
parametrization. The conventional expansion~\cite{Fong:1990nt} is extended by
higher-order terms in order to reproduce the full momentum interval
used here. The parameters of the distorted Gaussian are the mean
multiplicity $N$, the mean peak position $\bar{\xi}$, the width or
dispersion $\sigma$, the skewness $s$, and the kurtosis $k$. The
dependence of these fitted parameters on $X_b$ is shown separately in
Fig.~\ref{fig:xi_fit_parameters}. The parameters exhibit
an approximately linear dependence on $X_b$. Extrapolation to $X_b=0$
therefore yields a reconstructed parton spectrum in the limit where
the decay-dipole contribution is strongly suppressed.
The comparison with the independently generated parton momentum
spectrum in the stable-top jet, shown in
Fig.~\ref{fig:multiple_xi_partons}, provides the central closure
test of the reconstruction method. Around the maximum of the
distribution the difference is approximately $5\%$, and for
$\xi_p\gtrsim3$ it remains below about $12\%$. This agreement
indicates that the $\widehat{tb}$ contribution is effectively
removed; the residual difference provides an estimate of the
systematic uncertainty associated with the procedure.

The closure test also gives the stable-top reference a direct physical
role: the primary parton momentum distribution of a stable top-quark jet
can be reliably reconstructed from physical $t\to b\ell\nu$ events by
exploiting the angular dependence of the decay radiation.

This behaviour is markedly different from that observed for the
angular distributions in Fig.~\ref{fig:ang_dist_combined}. The distributions
for the unstable top quark in the region of small
$\Theta^2/\Theta_0^2$ do not converge towards the stable-top dead-cone
pattern as $X_b\to0$, whereas the inclusive momentum spectra exhibit a
smooth approach to the stable-top distribution over most of the
considered range. This contrast provides additional motivation for using the
momentum distribution as the quantitative observable for the
top-quark dead-cone effect.

\subsection{Dead cone in top-quark fragmentation at hadron level}
\label{subsec:top_fragmentation}

We finally return to the momentum-space formulation introduced for
charm and bottom at hadron level. For the top quark we apply the same MLLA estimate,
Eq.~(\ref{eq:MLLAeqxi}), now at $W=1$~TeV.
The \textsc{Pythia} simulation gives
\begin{equation}
    \langle x_t\rangle\simeq0.92,
    \qquad
    \xi_t\simeq0.083,
\end{equation}
while the corresponding mass-dependent scale is
\begin{equation}
    W_0=\sqrt{e}\,m_t
    \simeq286~{\rm GeV}.
\end{equation}
The much larger subtraction scale relative to charm and bottom
reflects the large top-quark mass and leads to a correspondingly
strong modification of the momentum spectrum
\cite{Dokshitzer:1991fc,Dokshitzer:1991fd}. This is verified experimentally at LEP energies for charm and bottom quarks and awaits verification at higher energies and for top quarks.


\begin{figure}[!t]
    \centering
    \begin{tabular}{cc}
    \includegraphics[width=0.92\columnwidth]
        {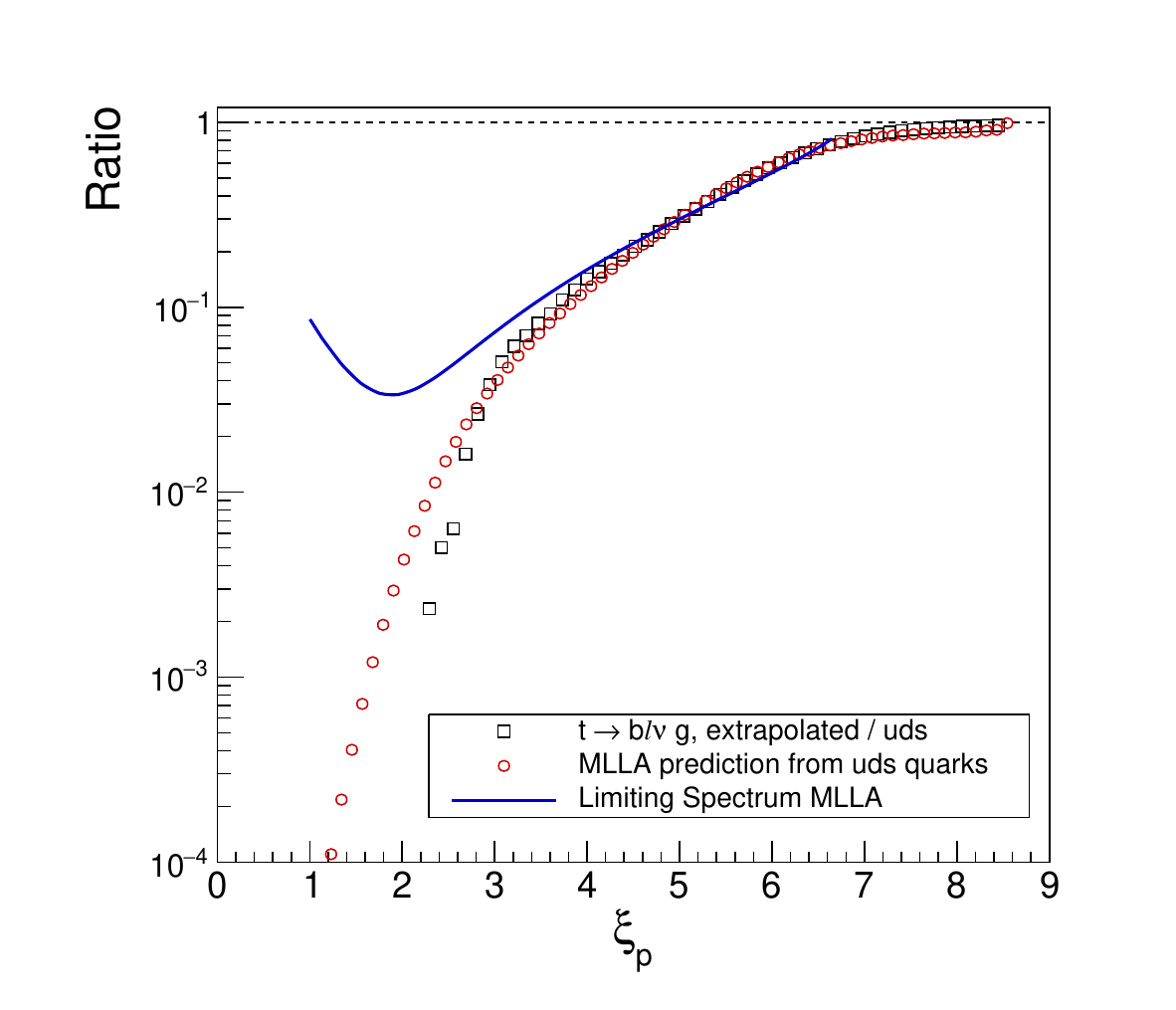}
    \end{tabular}
    \caption{
    Ratio of the top-quark fragmentation results to the corresponding
    light $uds$-quark fragmentation function at $W=1$~TeV, compared
    with the MLLA-based expectations. The figure illustrates the strong
    suppression towards small $\xi_p$, corresponding to large particle
    momenta, which reaches approximately two orders of magnitude in
    the high-momentum region (based on \textsc{Pythia} MC).
    }
    \label{fig:top_light_ratio}
\end{figure}

As shown in Fig.~\ref{fig:top_light_ratio}, the mass effect is expected to become
dramatic in momentum space. At large $\xi_p$, where soft particles
dominate, the heavy- and light-quark distributions approach one
another, whereas at small $\xi_p$ the ratio $t/uds$ is strongly
suppressed by two orders of magnitude and at $\xi_p\sim2$ it reaches
\begin{equation}
    \frac{\bar D_t}{\bar D_{uds}}
    \sim10^{-2}.
\end{equation}
%

\begin{figure}[!t]
    \centering
    \begin{tabular}{cc}
    \includegraphics[width=0.47\textwidth]
        {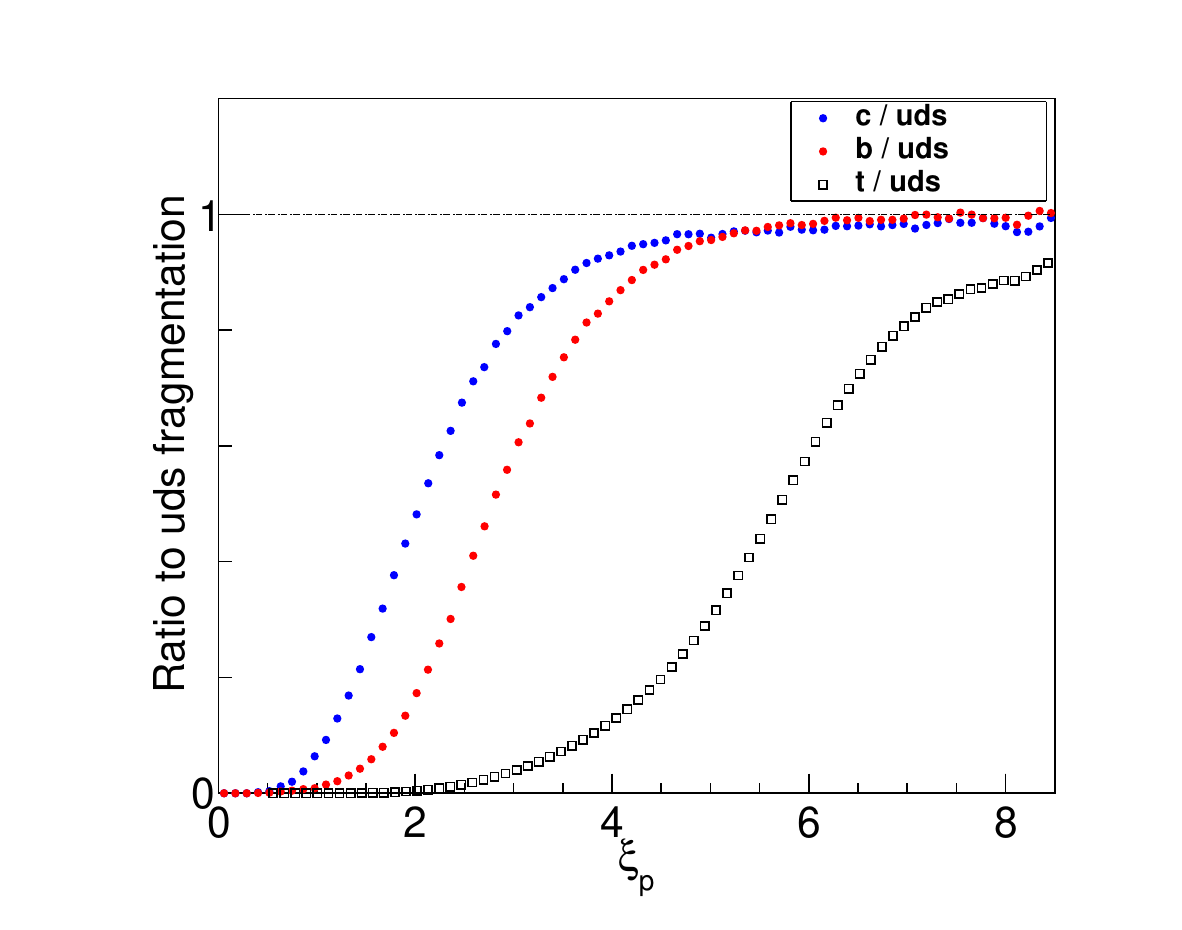}
    \end{tabular}
    \caption{
    Unified comparison of the fragmentation functions for charm,
    bottom, and top quarks at $\sqrt{s}=1$~TeV. The
    panel shows the ratios of the heavy-quark distributions to the
    light-quark reference. The progressive suppression towards small
    $\xi_p$, corresponding to large particle momenta, displays the
    strong mass dependence of the dead-cone effect (based on \textsc{Pythia} MC).  
    }
    \label{fig:deadcone_cbt}
\end{figure}

This result extends the trend already observed for charm and bottom
at LEP. The single-inclusive momentum distribution therefore provides
a particularly sensitive observable for following the mass dependence
of the dead cone up to the top-quark scale.
The MLLA description reproduces the reconstructed parton spectrum
associated with the top-quark jet over the central region at the level
of roughly $10$--$15\%$ in the \textsc{Pythia} MC.
As in the charm- and bottom-quark case, the comparison should not be
pushed to the extreme high-momentum boundary, where the approximations
underlying the MLLA differential-spectrum relation are not expected
to remain quantitatively reliable. The expectations for these ratios for the three heavy quarks, based on
the \textsc{Pythia} event generator, are shown in
Fig.~\ref{fig:deadcone_cbt}. They show the progressive suppression with
increasing quark mass.


\section{Conclusions}
\label{sec:conclusions}

We have investigated the dead-cone effect over a wide range of
heavy-quark masses 
combining its
geometric manifestation in angular radiation patterns with its
quantitative imprint on the single-particle inclusive momentum
distribution. The resulting picture provides a unified description
of the mass dependence of QCD radiation from the charm- and
bottom-quark scales up to the top-quark scale.

For charm and bottom quarks at the $Z$ pole, the characteristic
mass-dependent angular scales $\Theta_0=M_Q/E_Q$ provide the geometric basis of the
dead-cone picture. The parton-level bottom-quark radiation pattern in Fig.~\ref{fig:deadcone_b_LEP}
illustrates this depletion explicitly in the normalized $(X,Y)$ plane.
The same representation applied to a stable top quark at
$\sqrt{s}=1$~TeV in Fig.~\ref{fig:deadcone_stable_top_1TeV} reveals a similar structure after rescaling the angular
variables by the corresponding dead-cone angle $\Theta_0$.
The fragmentation functions in momentum space
reconstructed from LEP data provide a quantitative manifestation of the
same phenomenon. Compared to light-quark 
$uds$ jets, heavy-quark spectra are depleted at small $\xi_p=\ln(1/x_p)$,
corresponding to large particle momenta. The magnitude and flavour
dependence of this experimentally observed suppression are described by
the MLLA estimate over the central kinematic regions.

The extension to $\sqrt{s}=1$~TeV allows charm, bottom, and top quarks
to be compared within a common kinematic framework, which we explore
using \textsc{Pythia} Monte Carlo. For
$E_Q\simeq500$~GeV, their characteristic dead-cone angles satisfy
\begin{equation}
    \Theta_0^c \ll \Theta_0^b \ll \Theta_0^t,
\end{equation}
directly reflecting the mass hierarchy $m_c\ll m_b\ll m_t$.
As summarized in
Fig.~\ref{fig:deadcone_cbt}, the suppression relative to light-quark
fragmentation becomes progressively stronger from charm to bottom and
to top, reaching approximately two orders of magnitude for the top
quark in the high-momentum region in our top-quark study.

For the physical, unstable top quark, the primary dead-cone pattern is not
observed in isolation. The decay $t\to bW$ introduces additional
coherent radiation associated with the $\widehat{tb}$ colour dipole,
which partially populates the region primarily depleted.
This additional contribution is expected to vanish
if the $b$-quark direction approaches the
top-quark direction ($X_b\to0$) according to a calculation in lowest perturbative order.
The MC calculation shows indeed that the momentum distributions determined at a finite angle $X_b$ extrapolate towards a distribution at $X_b=0$, which is in close agreement with the momentum spectrum generated independently for the stable top quark.
The agreement within approximately $5\%$ near its maximum and approximately $12\%$
for $\xi_p\gtrsim3$ represents a non-trivial closure test of the
procedure, see Figs. \ref{fig:multiple_xi_partons} and \ref{fig:xi_fit_parameters}.

In pursuing a similar procedure for the angular distribution of gluons and partons around the top direction, we found that an extrapolation of these angular distributions in the b-quark angle with $X_b\to 0$ does not converge towards the corresponding angular distribution expected for the stable top-quark with the forward dip. The reason is that in the region of small emission angles $\Theta\sim 0$ around the top quark there is still a large contribution from primary gluons and partons emitted from the b-quark which enters the region close to top-quark direction overwhelming the low production rate of partons related to the top quark, according to the MC results.


The angular and momentum analyses are therefore complementary methods in the study of the dead cone:
the former provides a geometrical view at the radiation emitted from the heavy quark showing the depletion at small angles,
whereas the latter yields a highly sensitive quantitative measure of
the mass-dependent suppression of the high momenta. An advantage of the momentum analysis is the low sensitivity to the determination of the primary heavy quark direction and for the top-quark the possibility to isolate the primary top quark radiation by the extrapolation procedure in b-quark angle.

The combined charm, bottom, and top results reveal a continuous
evolution of mass-dependent QCD radiation across widely separated quark
masses. These results have already been confirmed experimentally for charm
and bottom quarks at LEP energies, while the extension to the top-quark
scale constitutes a prediction for a pronounced manifestation of the dead cone effect to be tested experimentally. 
\bibliographystyle{elsarticle-num}
\bibliography{hbpdeadcone}

\begin{thebibliography}{10}
\expandafter\ifx\csname url\endcsname\relax
  \def\url#1{\texttt{#1}}\fi
\expandafter\ifx\csname urlprefix\endcsname\relax\def\urlprefix{URL }\fi
\expandafter\ifx\csname href\endcsname\relax
  \def\href#1#2{#2} \def\path#1{#1}\fi

\bibitem{Kluth:2025xpj}
S.~Kluth, W.~Ochs, R.~Perez-Ramos, {How to identify the dead cone in the
  top-quark jet}, Phys. Rev. D 113~(11) (2026) 114048.
\newblock \href {http://arxiv.org/abs/2512.19874} {\path{arXiv:2512.19874}},
  \href {https://doi.org/10.1103/ckrs-wt2x} {\path{doi:10.1103/ckrs-wt2x}}.

\bibitem{Dokshitzer:1991fc}
Y.~L. Dokshitzer, V.~A. Khoze, S.~I. Troian, {Particle spectra in light and
  heavy quark jets}, J. Phys. G 17 (1991) 1481--1492.
\newblock \href {https://doi.org/10.1088/0954-3899/17/10/003}
  {\path{doi:10.1088/0954-3899/17/10/003}}.

\bibitem{Dokshitzer:1991fd}
Y.~L. Dokshitzer, V.~A. Khoze, S.~I. Troian, {On specific QCD properties of
  heavy quark fragmentation ('dead cone')}, J. Phys. G 17 (1991) 1602--1604.
\newblock \href {https://doi.org/10.1088/0954-3899/17/10/023}
  {\path{doi:10.1088/0954-3899/17/10/023}}.

\bibitem{Ermolaev:1981cm}
B.~I. Ermolaev, V.~S. Fadin, Log--log asymptotic form of exclusive
  cross-sections in quantum chromodynamics, JETP Lett. 33 (1981) 269--272.

\bibitem{Mueller:1981ex}
A.~H. Mueller, On the multiplicity of hadrons in {QCD} jets, Phys. Lett. B 104
  (1981) 161--164.
\newblock \href {https://doi.org/10.1016/0370-2693(81)90581-5}
  {\path{doi:10.1016/0370-2693(81)90581-5}}.

\bibitem{Dokshitzer1991Basics}
Y.~L. Dokshitzer, V.~A. Khoze, A.~H. Mueller, S.~I. Troyan, Basics of
  Perturbative QCD, Editions Frontières, Gif-sur-Yvette, France, 1991.

\bibitem{Kluth:2023umf}
S.~Kluth, W.~Ochs, R.~Perez~Ramos, {Observation of the dead cone effect in
  charm and bottom quark jets and its QCD explanation}, Phys. Rev. D 107~(9)
  (2023) 094039.
\newblock \href {http://arxiv.org/abs/2303.13343} {\path{arXiv:2303.13343}},
  \href {https://doi.org/10.1103/PhysRevD.107.094039}
  {\path{doi:10.1103/PhysRevD.107.094039}}.

\bibitem{DELPHI:1998cgx}
{DELPHI Collaboration}, {$\pi^\pm$, $K^\pm$, p and anti-p production in Z0
  ---\ensuremath{>} q anti-q, Z0 ---\ensuremath{>} b anti-b, Z$^0$
  ---\ensuremath{>} u anti-u, d anti-d, s anti-s}, Eur. Phys. J. C 5 (1998)
  585--620.
\newblock \href {https://doi.org/10.1007/s100529800989}
  {\path{doi:10.1007/s100529800989}}.

\bibitem{OPAL:1998arz}
{OPAL Collaboration}, {Measurements of flavor dependent fragmentation functions
  in Z0 --\ensuremath{>} q anti-q events}, Eur. Phys. J. C 7 (1999) 369--381.
\newblock \href {http://arxiv.org/abs/hep-ex/9807004}
  {\path{arXiv:hep-ex/9807004}}, \href {https://doi.org/10.1007/s100529901067}
  {\path{doi:10.1007/s100529901067}}.

\bibitem{ALICE:2021aqk}
{ALICE Collaboration}, Direct observation of the dead-cone effect in quantum
  chromodynamics, Nature 605~(7910) (2022) 440--446, erratum: Nature
  \textbf{607} (2022) no.7920, E22.
\newblock \href {http://arxiv.org/abs/2106.05713} {\path{arXiv:2106.05713}},
  \href {https://doi.org/10.1038/s41586-022-04572-w}
  {\path{doi:10.1038/s41586-022-04572-w}}.

\bibitem{LHCb:2025mcq}
{LHCb Collaboration}, {Measurement of the Lund plane for light- and
  beauty-quark jets}, Phys. Rev. D 112~(7) (2025) 072015.
\newblock \href {http://arxiv.org/abs/2505.23530} {\path{arXiv:2505.23530}},
  \href {https://doi.org/10.1103/r16d-b4my} {\path{doi:10.1103/r16d-b4my}}.

\bibitem{LHCb:2026mbl}
{LHCb Collaboration}, {B}-jet fragmentation with {{B}$^{\pm}\to $J$/\psi
  $K$^{\pm}$} decays in {$\sqrt{s}=13$ {TeV}} {$pp$} collisions at {LHCb},
  Phys. Rev. D 113~(11) (2026) 112006.
\newblock \href {http://arxiv.org/abs/2603.23663} {\path{arXiv:2603.23663}},
  \href {https://doi.org/10.1103/3d7r-c3y4} {\path{doi:10.1103/3d7r-c3y4}}.

\bibitem{CMS:2026lkj}
{CMS Collaboration}, Evidence for the dead-cone effect in bottom quark
  initiated jets produced in proton-proton collisions at $\sqrt{s}=13$
  {{{TeV}}} (2026).
\newblock \href {http://arxiv.org/abs/2608.29412} {\path{arXiv:2608.29412}}.

\bibitem{CMS:2025afq}
{CMS Collaboration}, {Exploring small-angle emissions in charm quark jets in
  proton-proton collisions at $\sqrt{s}=5.02$ {{TeV}}}, JHEP 05 (2026) 176.
\newblock \href {http://arxiv.org/abs/2507.13469} {\path{arXiv:2507.13469}},
  \href {https://doi.org/10.1007/JHEP05(2026)176}
  {\path{doi:10.1007/JHEP05(2026)176}}.

\bibitem{Bierlich:2022pfr}
C.~Bierlich, et~al., {A comprehensive guide to the physics and usage of PYTHIA
  8.3}, SciPost Phys. Codebases 8 (3 2022).
\newblock \href {http://arxiv.org/abs/2203.11601} {\path{arXiv:2203.11601}},
  \href {https://doi.org/10.21468/SciPostPhysCodeb.8}
  {\path{doi:10.21468/SciPostPhysCodeb.8}}.

\bibitem{Maltoni:2016ays}
F.~Maltoni, M.~Selvaggi, J.~Thaler, {Exposing the dead cone effect with jet
  substructure techniques}, Phys. Rev. D 94~(5) (2016) 054015.
\newblock \href {http://arxiv.org/abs/1606.03449} {\path{arXiv:1606.03449}},
  \href {https://doi.org/10.1103/PhysRevD.94.054015}
  {\path{doi:10.1103/PhysRevD.94.054015}}.

\bibitem{Cacciari:2011ma}
M.~Cacciari, G.~P. Salam, G.~Soyez, Fastjet user manual, Eur. Phys. J. C 72
  (2012) 1896.
\newblock \href {http://arxiv.org/abs/1111.6097} {\path{arXiv:1111.6097}},
  \href {https://doi.org/10.1140/epjc/s10052-012-1896-2}
  {\path{doi:10.1140/epjc/s10052-012-1896-2}}.

\bibitem{Cacciari:2005hq}
M.~Cacciari, G.~P. Salam, Dispelling the n\^(3) myth for the k\_t jet-finder,
  Phys. Lett. B 641 (2006) 57--61.
\newblock \href {http://arxiv.org/abs/hep-ph/0512210}
  {\path{arXiv:hep-ph/0512210}}, \href
  {https://doi.org/10.1016/j.physletb.2006.08.037}
  {\path{doi:10.1016/j.physletb.2006.08.037}}.

\bibitem{Dokshitzer:2005ri}
Y.~L. Dokshitzer, F.~Fabbri, V.~A. Khoze, W.~Ochs, {Multiplicity difference
  between heavy and light quark jets revisited}, Eur. Phys. J. C 45 (2006)
  387--400.
\newblock \href {http://arxiv.org/abs/hep-ph/0508074}
  {\path{arXiv:hep-ph/0508074}}, \href
  {https://doi.org/10.1140/epjc/s2005-02424-5}
  {\path{doi:10.1140/epjc/s2005-02424-5}}.

\bibitem{DELPHI:2011aa}
J.~Abdallah, et~al., {A study of the b-quark fragmentation function with the
  DELPHI detector at LEP I and an averaged distribution obtained at the Z
  Pole}, Eur. Phys. J. C 71 (2011) 1557.
\newblock \href {http://arxiv.org/abs/1102.4748} {\path{arXiv:1102.4748}},
  \href {https://doi.org/10.1140/epjc/s10052-011-1557-x}
  {\path{doi:10.1140/epjc/s10052-011-1557-x}}.

\bibitem{Baines:2006uw}
J.~Baines, et~al., {Heavy quarks (Working Group 3): Summary Report for the
  HERA-LHC Workshop Proceedings} (1 2006).
\newblock \href {http://arxiv.org/abs/hep-ph/0601164}
  {\path{arXiv:hep-ph/0601164}}.

\bibitem{ParticleDataGroup:2024cfk}
S.~Navas, et~al., Review of particle physics, Phys. Rev. D 110~(3) (2024)
  030001.
\newblock \href {https://doi.org/10.1103/PhysRevD.110.030001}
  {\path{doi:10.1103/PhysRevD.110.030001}}.

\bibitem{Orr:1992uv}
V.~A. Khoze, W.~J. Stirling, L.~H. Orr, Soft gluon radiation in $e^{+} e^{-}
  \to t \bar{t}$, Nucl. Phys. B 378 (1992) 413--442.
\newblock \href {https://doi.org/10.1016/0550-3213(92)90603-9}
  {\path{doi:10.1016/0550-3213(92)90603-9}}.

\bibitem{Jikia:1991yd}
G.~V. Jikia, Measuring the heavy top quark width by soft gluon radiation in
  $e^{+}e^{-} \to t\bar{t}$, Phys. Lett. B 257 (1991) 196--200.
\newblock \href {https://doi.org/10.1016/0370-2693(91)90881-P}
  {\path{doi:10.1016/0370-2693(91)90881-P}}.

\bibitem{Fong:1990nt}
C.~P. Fong, B.~R. Webber, One and two particle distributions at small {$x$} in
  {QCD} jets, Nucl. Phys. B 355 (1991) 54--81.
\newblock \href {https://doi.org/10.1016/0550-3213(91)90302-E}
  {\path{doi:10.1016/0550-3213(91)90302-E}}.

\end{thebibliography}


\end{document}